# An Overview of Exoplanet Biosignatures

Edward W. Schwieterman and Michaela Leung

*Department of Earth and Planetary Sciences*
*University of California, Riverside*
*Riverside, CA, 92521, U.S.A.*
*eschwiet@ucr.edu, mleun019@ucr.edu*

## INTRODUCTION: THE SEARCH FOR LIFE BEYOND THE SOLAR SYSTEM

### Habitable exoplanets and the context for biosignatures

In the last three decades, knowledge of planetary systems other than our own has increased rapidly with the discovery of over 5,000 exoplanets (Christiansen 2022). These worlds vary greatly in mass, composition, and insolation, encompassing planets ranging in size from smaller than Mercury (Barclay et al. 2013) to more than twice the radius of Jupiter (Crouzet et al. 2017). Most recently, exoplanetary science has entered a new era of planetary characterization with the launch of the James Webb Space Telescope (JWST), which has already unveiled the composition and chemical processes of exoplanet atmospheres in unprecedented detail (e.g., Ahrer et al. 2023; Alderson et al. 2023; Tsai et al. 2023). The study of exoplanet atmospheres is advancing at an unprecedented rate, but current reviews can provide a foundational basis for understanding their formation, dynamics, climates, chemistries, and observables (Helling 2019; Jontof-Hutter 2019; Pierrehumbert and Hammond 2019; Shields 2019; Madhusudhan 2019; Wordsworth and Kreidberg 2022; Kempton and Knutson 2024).

A subset of discovered exoplanets orbit within the so-called "habitable zone" (HZ), most often defined as the range of distances from a star where a geologically active rocky planet with an $N_2$-$CO_2$-$H_2O$ atmosphere can maintain temperatures suitable for surface liquid water (Kasting et al. 1993; Kopparapu et al. 2013, 2014; Kane et al. 2016). The HZ can also be defined more broadly and may encompass super-Earths with $H_2$-dominated atmospheres and deep global oceans, called Hycean (**hy**drogen o**cean**) worlds (Madhusudhan et al. 2021, 2023). While liquid water is a key requirement for life as we know it, habitability as a concept encompasses other potential requirements, including the availability of energy, nutrients, and other hospitable physiochemical conditions beyond clement temperatures (Cockell et al. 2016; Hoehler et al. 2020).

The first step in searching for evidence of life is characterizing the habitability of the star-planet system. The HZ can predict the planetary targets most likely to host surface oceans, allowing for the robust exchange of gases between a potential biosphere and the atmosphere (Kasting et al. 2014) and potentially those targets for which access to information about the planetary surface is obtainable remotely. The HZ as a conceptual tool is intended to guide the design of telescopes and surveys to find habitable worlds with remotely characterizable atmospheres and surfaces and does not (and was never intended to) circumscribe all—or possibly even most—environments in which life could arise or is currently present. For example, icy moons, also termed ocean worlds, such as analogs to our solar system moons Europa and Enceladus, may be common habitable environments in the universe (Hendrix et al. 2019). Still, their potentially habitable subsurface environments are inaccessible for remote characterization.

The search for life beyond the solar system is a central goal of the NASA Astrobiology Program (Hays et al. 2015) and one of the driving motivations behind the National Academy of Sciences's recommendations for a proposed infrared (IR) / optical / Ultraviolet (UV) Surveyor to succeed JWST and the Hubble Space Telescope (National Academies of Sciences, Engineering, and Medicine 2021). Such evidence for life, termed "biosignatures," could be determined from the spectral characterization of these distant worlds.

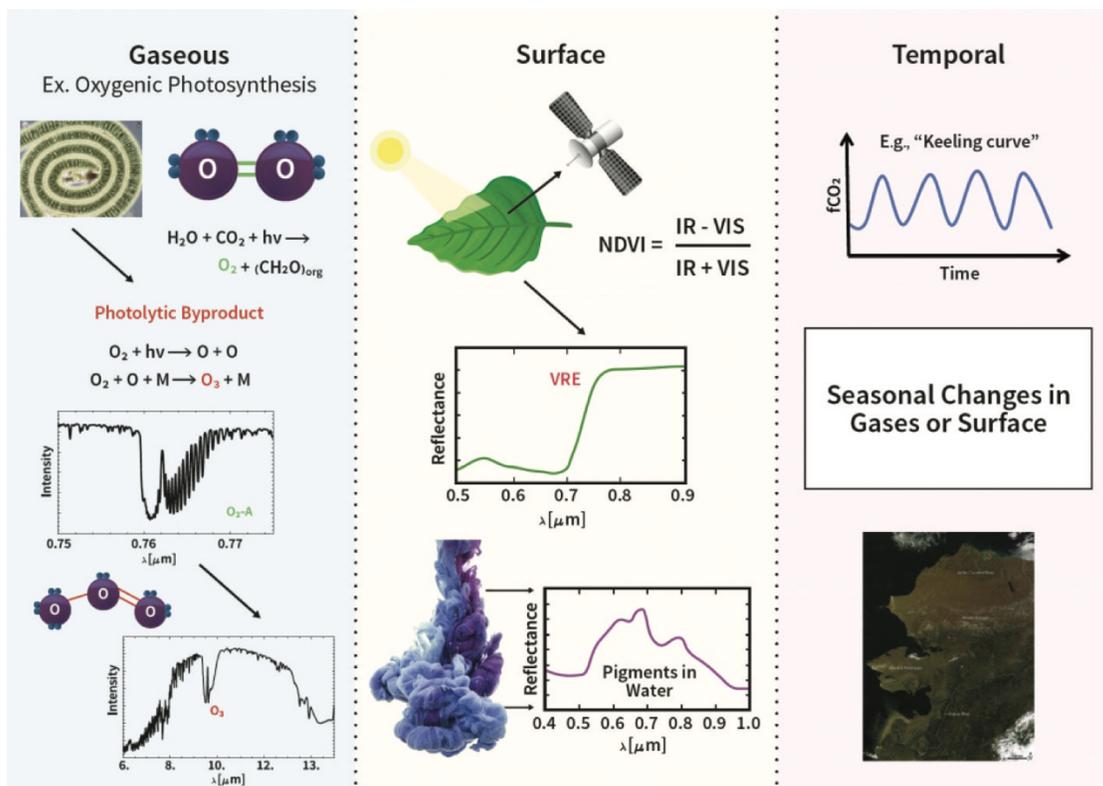

**Figure 1.** Summary of gaseous (**left**), surface (**center**), and temporal (**right**) biosignatures. Gaseous biosignatures include spectrally active, volatile metabolic products, such as $O_2$ produced by oxygenic photosynthesis, and potential photochemical byproducts, such as ozone ($O_3$) from $O_2$ photochemistry. Surface biosignatures include the vegetation red edge (VRE), which results from the sharp contrast between chlorophyll absorption at visible wavelengths and scattering at infrared wavelengths in photosynthetic organisms. Temporal biosignatures include time-dependent modulation of gases linked to life—such as seasonal changes in $CO_2$ consumed by photosynthesis and released by the decay of organic materials—or variations in albedo from the growth and decay of vegetation. This figure is reproduced from Schwieterman (2021) under Creative Commons Attribution License CC-BY. Sub image credits: NASA and the Encyclopedia of Life (EOL).

**What is an exoplanet biosignature?**

A biosignature is generally defined as "substances, structures, patterns, or processes, or ensembles of these features" that indicate the current or former presence of life and can be separated from abiotic sources (Des Marais and Walter 1999; Des Marais et al. 2008; Hays et al. 2015). In the context of exoplanets, a biosignature is a remotely observable indication of living processes influencing a planet's atmosphere or surface. A biosignature can be a gaseous molecule (or suite of molecules), a surface feature (or suite of features), or time-dependent modulations of gases or surface features that can be linked to life (Figure 1). We narrow this definition to exclude signs of technological life, called technosignatures (Tarter 2001, 2006; Wright et al. 2022), which are beyond the scope of this chapter. (However, see Haqq-Misra et al. (2022) for an overview of potential planetary technosignatures).

Any purported exoplanet biosignature will require further vetting beyond the first observation to confirm a biological origin. So, it is best to designate all exoplanet biosignatures as "potential biosignatures" until such vetting has brought reasonable certainty to that designation (Schwieterman et al. 2018a; Meadows et al. 2022). We note in passing that some authors have expressed skepticism that any remote signature could robustly indicate (non-technological-)life (e.g., Smith and Mathis 2023) or take issue with the term "biosignature" as typically applied to potential signs of life on exoplanets and seek to redefine it (e.g., Gillen et al. 2023). While these perspectives are important and should be discussed (Malaterre et al. 2023), we will not litigate them here, and our remit is to provide an overview of exoplanet biosignatures as proposed in the literature.

A compelling biosignature has the following attributes: detectability, survivability, and specificity (Meadows 2017; Meadows et al. 2018b). Detectability refers to the intrinsic absorption, scattering, or emission properties of the molecule or feature. To be observed, a biosignature must influence the information embedded within light that is reflected or emitted from or transmitted through the target body. For example, the $O_2$ molecule has a strong absorption feature at 0.76 μm that is apparent in the reflected light spectrum of Earth (Figure 2). Survivability refers to the biosignature molecule or substance's ability to endure rapid depletion or destruction by planetary or astrophysical factors such as the host star's ultraviolet (UV) radiation. A useful biosignature must have the capacity to accumulate to levels that can meaningfully impact the planet's spectrum, an outcome controlled by factors such as chemical kinetics and dissociation cross-sections in the context of atmospheric biosignatures. However, robustness to environmental depletion can cut both ways if small abiotic sources of potential biosignature molecules can accumulate over long timescales. Finally, specificity refers to the separability of biotic and abiotic sources, which will always depend on the available context. Earth's $CH_4$, for example, is in a strong kinetic and thermodynamic disequilibrium with our planet's temperate $O_2$-rich atmosphere (Sagan et al. 1993). In contrast, methane is an abundant gas in Jupiter, Saturn, Uranus, and Neptune, as it is an expected equilibrium product in massive, primordial $H_2$-rich atmospheres with high-temperature interiors (Lodders and Fegley 2002; Moses et al. 2013).

Here, we provide an overview of proposed exoplanet biosignatures, including their biological origins, observable features, and potentially confounding abiotic sources. Additional information is available within the cited sources herein and past comprehensive reviews (Seager et al. 2012; Seager and Bains 2015; Kaltenegger 2017; Grenfell 2017; Schwieterman et al. 2018a; Catling et al. 2018; Fujii et al. 2018; Walker et al. 2018). We aim to emphasize material published since these reviews while providing a foundational understanding of each named biosignature.

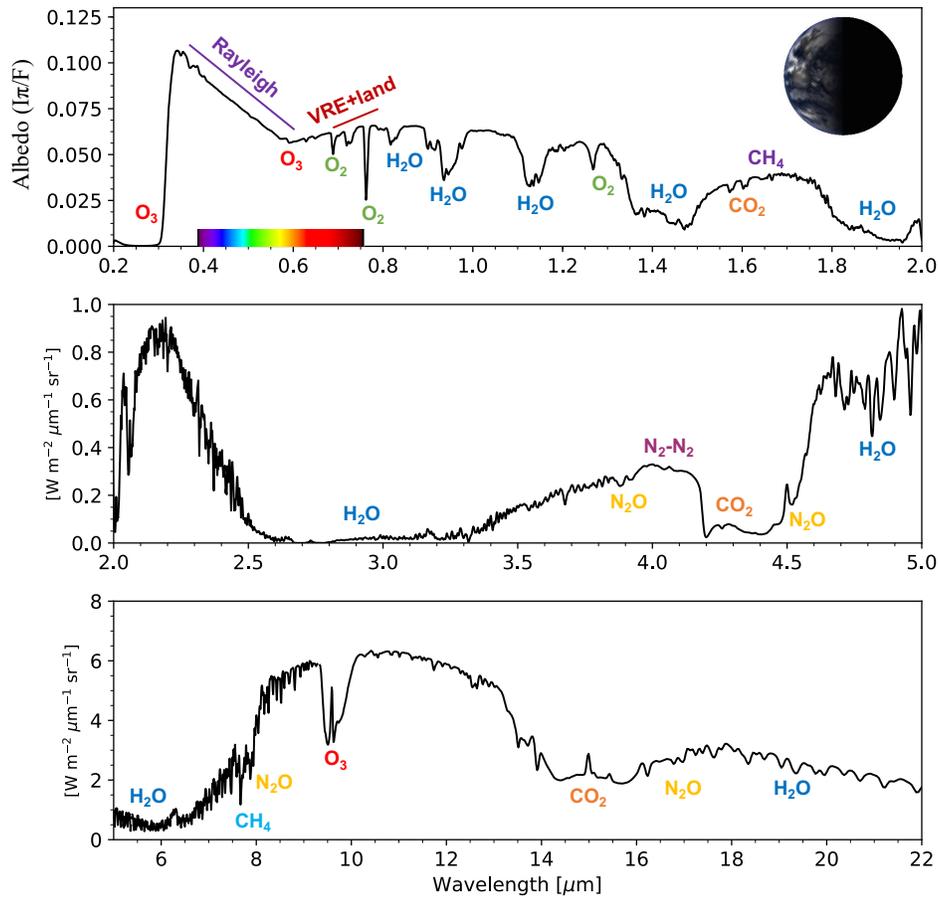

**Figure 2.** Simulated spectrum of Earth from 0.2-22 μm at quadrature-phase (half illumination) showing various spectral features that include biosignature gases $O_2$, $O_3$, and $CH_4$; the vegetation red edge (VRE) surface feature; and habitability marker gases $H_2O$, $CO_2$, and $N_2$. Rayleigh scattering is indicative of atmospheric pressure. **Top:** Apparent spectral albedo in reflected light from 0.2-2 μm (ultraviolet/visible/near-infrared) at quadrature phase. **Middle:** Near-infrared (2-5 μm) spectral radiance (units: W m$^{-2}$ μm$^{-1}$ sr$^{-1}$), including reflected and emitted light components. **Bottom:** Thermal infrared (5-22 μm) spectral radiance. The synthetic spectrum was calculated using the Virtual Planetary Laboratory 3D spectral Earth model (Robinson et al. 2011; Schwieterman et al. 2015b). This figure is reproduced with minor modifications from Schwieterman et al. (2018a) under <u>Creative Commons Attribution License CC-BY</u>. Modifications include additional gas feature labels, an inset of a simulated Earth at half illumination, and an **inset color bar** indicating the visible wavelength range.

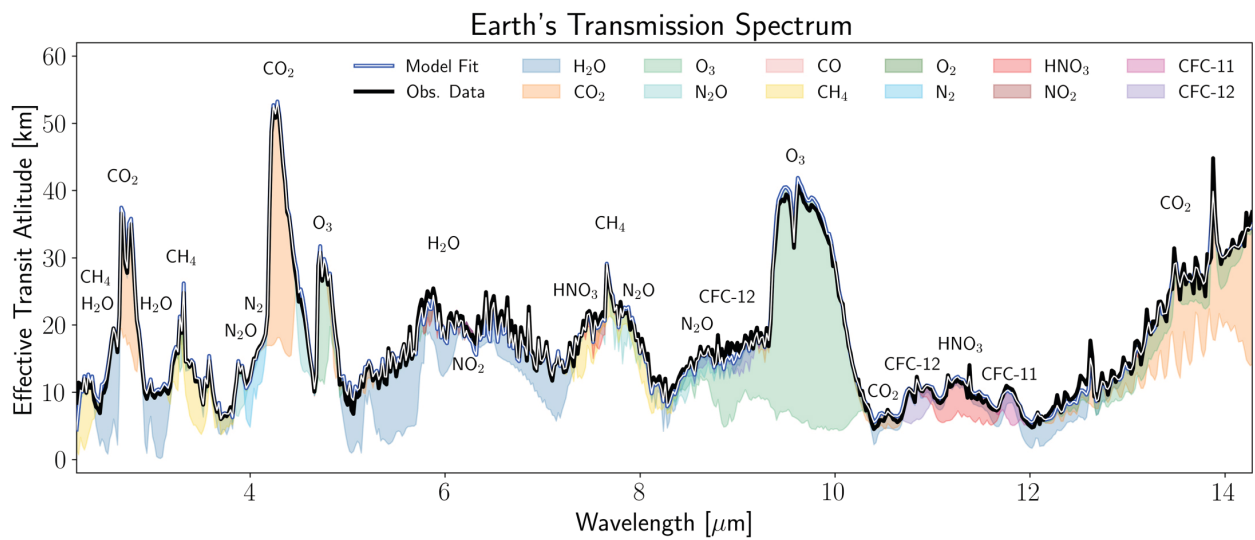

**Figure 3.** Earth's infrared (2-15 μm) cloud-free transmission spectrum from Macdonald and Cowan (2019; **black line**) and fit using the SMARTER retrieval model (**blue line**; Lustig-Yaeger et al. 2023). This spectrum shows gaseous habitability markers $CO_2$ (2.7, 4.3, 15 μm), $H_2O$ (6 μm), and $N_2$ (4.2 μm); biosignatures $CH_4$ (3.3, 7.7 μm), $O_3$ (4.7, 9.7 μm) and $N_2O$ (4, 7.7, 8.5 μm), and industrial pollutants CFC-11 ($CCl_3F$; 11.8 μm) and CFC-12 ($CCl_2F_2$; 10.8 μm), and $NO_2$ (6.2 μm). This figure is taken from Lustig-Yaeger et al. (2023) under <u>Creative Commons Attribution License CC-BY</u>.

## ATMOSPHERIC (GASEOUS) BIOSIGNATURES

Atmospheric, or gaseous, biosignatures are volatile molecules that are either direct products of life or secondary products from the environmental processing of biogenic compounds. Biogenic gas emissions can include those directly related to the primary energy-yielding metabolism (such as $O_2$ from oxygenic photosynthesis) or incidental products from other cellular processes. Biogenic substances such as DNA and RNA make poor remotely detectable biosignatures due to their high molecular weight, low volatility, significant fragility, and broad and ambiguous spectral features. Thus, the search for atmospheric biosignatures is limited to small, volatile molecules with a reasonable chance of accumulating to detectable concentrations. Because all small molecules have abiotic sources—though some are much more limited than others—the interpretation of any of these molecules as a sign of life highly depends on the context in which they are found (Krissansen-Totton et al. 2022). It is crucial to emphasize that planetary atmospheres and surfaces can consume molecules at high rates, so the mere existence of molecules in a non-planetary context, such as the interstellar medium (e.g., McGuire 2018), does not (necessarily) negate their potential utility for fingerprinting life in the proper context.

Gaseous biosignatures may be observed via transmitted, reflected, or emitted light from a planetary atmosphere (Figs. 2, 3). Therefore, biosignature gases must interact with light via dissociation, electronic, or vibrational transitions to be spectrally observable. Figure 4 shows the intrinsic near-to-far infrared (IR) opacities for several proposed biosignature molecules with data sourced primarily from the 2020 HITRAN database and original data sources (Sharpe et al. 2004; Gordon et al. 2022). Tables 1 and 2 summarize several key gaseous biosignatures, including their modern concentration on Earth, the environments and organisms that produce them, the wavelengths of their major absorption features, and their known abiotic sources. Below, we describe each gas in more detail, including its sources, sinks, and potentially observable features.

## Oxygen and Ozone ($O_2$ and $O_3$)

*Oxygen ($O_2$).* The most referenced gaseous biosignature is molecular oxygen ($O_2$), a product of oxygenic photosynthesis on Earth. Photosynthesis is a means by which life can convert light energy, inorganic carbon in the form of carbon dioxide ($CO_2$), and water into organic biomass. The net equation for oxygenic photosynthesis can be given as:

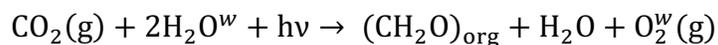

$$CO_2(g) + 2H_2O^w + h\nu \rightarrow (CH_2O)_{org} + H_2O + O_2^w(g)$$

where $h\nu$ is the energy of the photon(s) involved in the reaction (where h is Planck's constant and $\nu$ is the photon's frequency) and $(CH_2O)_{org}$ represents organic matter (biomass). Here we have used a *w* superscript to indicate that the $O_2$ resulting from the net reaction has been sourced from $H_2O$ rather than $CO_2$. We can generally describe oxygenic photosynthesis as a reduction-oxidation reaction where electrons are transferred from $H_2O$, the reductant, through complex biochemical reactions, to carbon dioxide, thus reducing the carbon and evolving a waste oxidant ($O_2$). Notably, the simplified net reaction presented here belies significant metabolic complexity beyond the scope of this chapter (see, e.g., Hohmann-Marriott and Blankenship 2012 for a detailed treatment). Oxygenic photosynthesis is an immensely productive metabolism because it leverages common molecules $H_2O$ and $CO_2$ and bountiful photons from the Sun. Other forms of photosynthesis exist but are limited by reductants less common on terrestrial planets, such as $H_2$, $H_2S$, and ferrous iron ($Fe^{2+}$). (On an $H_2$-rich planet, $CO_2$ may instead be limiting). Therefore, oxygenic photosynthesis may be the most productive metabolism on any planet orbiting a star (Kiang et al., 2007a,b).

Oxygenic photosynthesis relies on Photosynthetically Active Radiation (PAR) photons in the ~0.4-0.72 μm range, while anoxygenic photosynthesis can use longer wavelength, lower energy photons into the near-infrared (NIR; Hohmann-Marriott and Blankenship 2012). The long-wavelength limit for oxygenic photosynthesis depends on many factors, including the primary photosynthetic chlorophyll pigments, but practical limits such as photon flux may mean oxygenic photosynthesis could be light-limited around cool M dwarf stars (Lehmer et al. 2018).

Importantly, the accumulation of oxygen in Earth's atmosphere involves complex long-term biogeochemical cycles and requires the burial of organic carbon to prevent the back reaction of the equation above for net oxygenation to occur (Catling 2014). Without resupply from photosynthesis (and organic carbon burial), the oxygen in Earth's atmosphere would be depleted by oxidative weathering and reactions with reduced volcanic gases over geologically short timescales (Lécuyer and Ricard 1999). See Stüeken et al. (2024), Meadows et al. (2018), and Lyons et al. (2014, 2021) for more detailed histories of the oxidation and oxygenation of Earth's atmosphere and accompanying implications for the remote detectability of oxygenic photosynthesis over geologic time.

It has long been recognized that the near-infrared and optical spectral bands of oxygen at 0.76 um (Fraunhoffer-A) and 0.69 μm (Fraunhoffer-B) could provide remote indications of Earth's atmospheric oxygen and, by extension, its photosynthetic biosphere (Owen 1980; Sagan et al. 1993). Oxygen is an interesting molecule as it is a symmetric, homonuclear molecule that nonetheless possesses significant electronic transitions in the optical and near-infrared spectrum, in contrast to $N_2$, which possesses few potentially detectable bands and only through collisionally induced absorption and dimer features in the infrared (Lafferty et al. 1996; Schwieterman et al. 2015b). Oxygen also absorbs meaningfully at 0.63 μm ($O_2$-γ) and 1.27 μm. In addition, there are $O_2$-$O_2$ collisionally induced absorption features (sometimes called $O_4$ dimer features, or $(O_2)_2$ features) at UV/Vis/NIR wavelengths at 0.345, 0.36, 0.38, 0.445, 0.475, 0.53, 0.57, 0.63, 1.06, and 1.27 μm (Greenblatt et al. 1990; Maté et al. 1999; Hermans et al. 1999; Karman et al. 2019) and in the mid-infrared (MIR) at 6.4 μm (Timofeev and Tonkov 1979; Maté et al. 2000; Fauchez et al. 2020).

***Detecting $O_2$.*** The observational manifestation of oxygen ($O_2$) biosignatures will depend on observing mode, wavelength coverage, and spectral resolving power. (This is generally true for all spectrally active gas species.) The 0.76 μm band is a commonly discussed target in reflected light due to its intrinsic strength, co-location with the high spectral fluxes from main sequence stars (and thus high planet flux in reflected light), and proximity to other absorption features of interest, such as water vapor. The Science Technology Definition Teams (STDT) for future reflected light direct-imaging mission proposals have used this band as a fiducial target in a terrestrial exoplanet atmosphere (The LUVOIR Team 2019; Gaudi et al. 2020).

Unfortunately, the $O_2$-A band's features are too narrow and weak to be detected by JWST via transit transmission observations at modern Earth-like abundances (Tremblay et al. 2020; Pidhorodetska et al. 2020). However, other molecules may be characterized in terrestrial atmospheres (Meadows et al. 2023). In some scenarios, abiotic $O_2$ may accumulate to substantially higher levels than found on modern Earth, which may be more detectable in transmission or reflected light spectra (see below). High-resolution (R~100,000) ground-based observations by Extremely Large Telescopes (ELTs) may also search for $O_2$ in exoplanetary atmospheres (Snellen et al. 2013). However, such searches present many challenges and may require decades or more observing time for a positive result (Currie et al. 2023; Hardegree-Ullman et al. 2023).

The most viable approach for detecting Earth-like levels of $O_2$ in exoplanetary atmospheres is via reflected light direct imaging with a space-based telescope, such as the Infrared/Optical/Ultraviolet (IR/O/UV) Surveyor recommended by the 2020 Astronomy & Astrophysics Decadal Survey (National Academies of Sciences, Engineering, and Medicine 2021). NASA's current designation for this mission concept is the Habitable Worlds Observatory (HWO;

), whose specific architecture and capabilities have yet to be determined.

   ***Ozone ($O_3$).*** Ozone is a photochemical by-product of atmospheric oxygen. Most of Earth's $O_3$ is generated in the stratosphere (15-50 km altitude) due to the photolysis of $O_2$. The dominant photochemical reactions that produce and consume $O_3$ in Earth's atmosphere are often called the "Chapman reactions" (Chapman 1930):

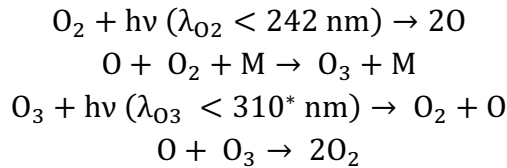

Ozone concentrations peak in Earth's stratosphere near 10 parts-per-million (ppm) at 30 km altitude, though both the peak concentration and the altitude at which that peak occurs vary spatially and temporally (Camp et al. 2003). The absorption of shortwave (UV/Visible) radiation by $O_3$ heats the upper atmosphere and creates a temperature inversion in Earth's stratosphere.

   The concentration of $O_3$ is strongly sensitive to the intensity and spectral energy distribution of the incident UV flux, and so predicted $O_3$ concentrations can be markedly different assuming non-Sun-like host stars and/or different flux-equivalent distances from those stars even for planets with identical atmospheric $O_2$ levels (Segura et al. 2003; Grenfell et al. 2007, 2014), and these predictions are very sensitive to the input stellar spectra (Cooke et al. 2023). In addition to UV radiation, stellar energic particles (StEP) from flare events can deplete $O_3$ columns, depending on the intensity and frequency of these events (Tabataba-Vakili et al. 2016; Tilley et al. 2019). There is a growing recognition that StEPs are an important area of study to fully understand the atmospheric chemistry of exoplanets (Airapetian et al. 2020; Chen et al. 2020; Garcia-Sage et al. 2023).

   Tropospheric (0-15 km) $O_3$ exists at substantially lower concentrations (~10-20 parts-per-billion, ppb) than in the stratosphere but is very important for tropospheric chemistry. The formation of $O_3$ is temperature-dependent (Burkholder et al. 2019; Pidhorodetska et al. 2021), and radical species, including OH, Cl, and Br, can catalytically destroy ozone (Solomon 1999), both factors that may further complicate predicting $O_3$ abundances from $O_2$ levels alone (Kozakis et al. 2022).

   Ozone possesses significant absorption features in the UV/Vis/NIR and mid-infrared wavelength regimes (**Figs. 2-4**). The strongest mid-infrared feature is the 9.65 μm, which has long been proposed as an indirect indicator of $O_2$ in the mid-infrared (e.g., Leger et al. 1993), where $O_2$ bands are non-existent or too weak to detect. The UV $O_3$ Hartley band is centered near 0.25 μm and extends from 0.2-0.31 μm; this band's absorption shields Earth's surface from deleterious UV radiation (ozone and other molecules such as $O_2$ and $CO_2$ absorb at shorter wavelengths). The weaker, structured Huggins UV band spans the range 0.31-0.36 μm. Ozone also absorbs in the visible Chappuis band between 0.4 and 0.7 μm with a peak near 0.6 μm. The weaker near-IR Wulf band extends from the long wavelength end of the Chappuis band to beyond 1 μm. Because $O_3$ absorption is continuous (though varying in opacity) throughout the UV/Vis/NIR, the literature varies concerning the wavelength cutoffs between each named region. $O_3$ has weaker features—that nonetheless vary greatly in strength—throughout the NIR and MIR, including at 2.05, 2.15, 2.5, 2.7, 3.3, 3.6, 4.7, 5.5, 5.8, 7.8, 8.9, and 14.0 μm (Gordon et al. 2022).

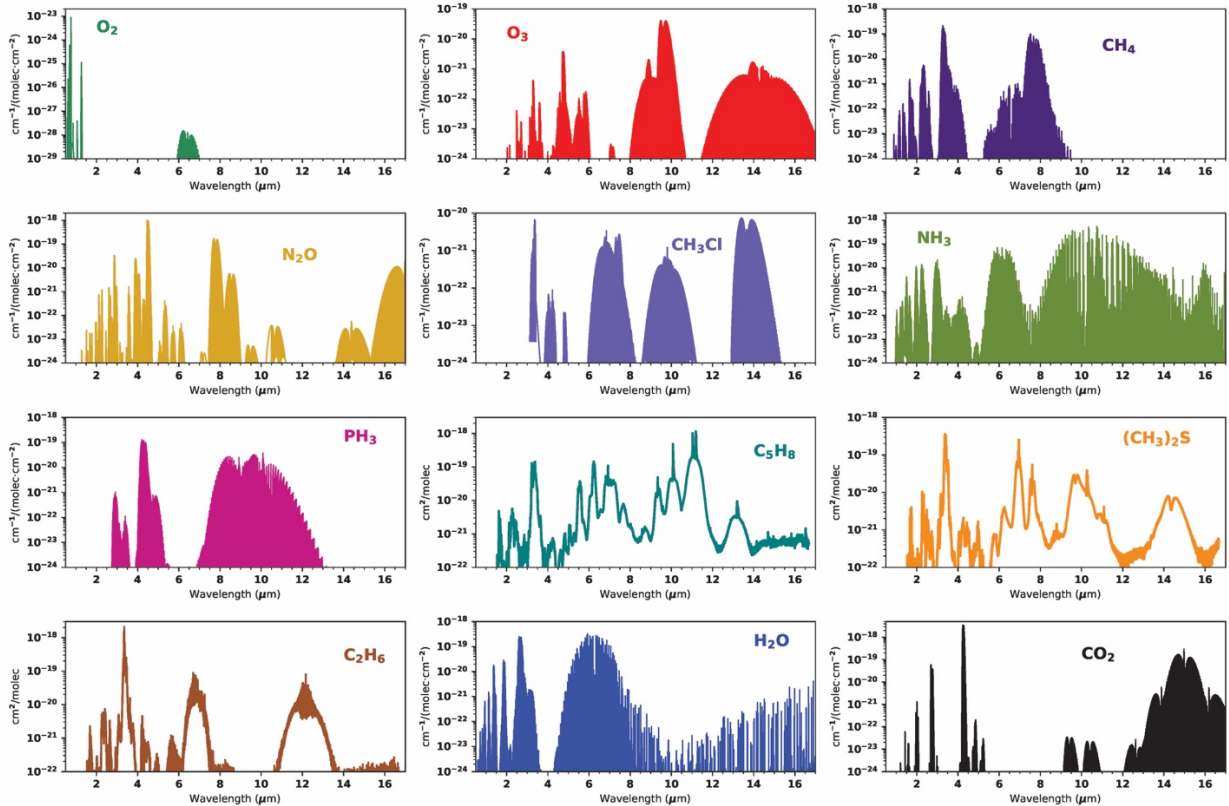

**Figure 4.** Absorption features for potential biosignature molecules and other gases important for terrestrial planet characterization. Shown are line intensities for the most abundant isotopologue from the HITRAN 2020 database (Gordon et al. 2022) for $O_2$, $O_3$, $CH_4$, $N_2O$, $CH_3Cl$, $NH_3$, $PH_3$, $H_2O$, and $CO_2$ in units of $cm^{-1}/(molec \cdot cm^{-2})$ and absorption cross-sections from the PNNL database (Sharpe et al. 2004) for $C_5H_8$, $(CH_3)_2S$ (DMS), and $C_2H_6$ in units of $cm^2/molecule$. Note that opacities are often incomplete for $\lambda < 2$ μm either because these data are not in HITRAN but are available elsewhere (e.g., $O_3$) or because they haven't been measured (e.g., $CH_3Cl$).

***Detecting $O_3$.*** Detecting ozone in the transmission spectra of temperate rocky exoplanets may be challenging. Existing studies find that over 100 transits with JWST's NIRSpec or MIRI instruments would be required to detect $O_3$ at $3\sigma$ significance on TRAPPIST-1e if it were an Earth-like planet (Lustig-Yaeger et al. 2019; Wunderlich et al. 2019, 2020; Gialluca et al. 2021), which is at or beyond the upper range of the transits that could be viewed in JWST's nominal five-year lifetime. These results depend on how the $O_3$ formation is modeled, which is critically dependent on the input stellar spectra, and studies of $O_3$ detectability that assume Earth's $O_3$ profile have found fewer transits would be required to detect $O_3$ on TRAPPIST-1e. For example, Lustig-Yaeger et al. (2023) find that the 4.7 μm $O_3$ feature can be constrained twice as precisely with NIRSpec than the 9.65 μm $O_3$ band can be constrained with MIRI. However, these results assume the actual Earth's spectrum rather than one based on self-consistent photochemistry with the host star. Notably, the detectability of $O_3$ (and other gases) in transmission spectroscopy is related to their intrinsic molecular opacities and the backlighting effect from the host star. Because the star is brighter in the NIR, features with lower intrinsic opacities can, therefore, be more detectable in practice. It is important to note that $O_3$ detectability in transmission spectra also depends on scale height, with higher temperatures allowing for greater transit signatures but, on the other hand, suppressing the chemical formation of $O_3$ (Pidhorodetska et al. 2021; Harman et al. 2022).

O$_3$ could be detected in reflected light by a future space-based UV/Vis/IR telescope such as HWO via its UV Hartley absorption at ~0.3 μm and broad visible absorption resulting from its Chappuis band (The LUVOIR Team 2019; Gaudi et al. 2020; Damiano et al. 2023). Additionally, the 9.65 μm $O_3$ band is one of the primary target molecules for missions designed to directly image planets in the MIR, such as ESA's Large Interferometer For Exoplanets (LIFE) mission concept (Konrad et al. 2022; Quanz et al. 2022; Alei et al. 2022). Because the formation of $O_3$ is non-linear with $O_2$ abundance, it could be a sensitive indicator of $O_2$ levels too low to detect directly (Reinhard et al. 2017, 2019; Schwieterman et al. 2018b; Olson et al. 2018), such as may have been the case during parts of Earth's Proterozoic Eon (0.5-2.5 Ga). However, caution is warranted given the dependence of $O_3$ formation and destruction on numerous other factors (Kozakis et al. 2022; Cooke et al. 2023), and the dependent of $O_3$ detectability in the IR on the temperature structure of the atmosphere.

**False Positives for Oxygen and Ozone Biosignatures**

Oxygenic photosynthesis produces essentially all the free $O_2$ in Earth's atmosphere today. Decades ago, there was a general view in the astrobiology community that detectable free $O_2$ was unlikely in the atmosphere of a habitable planet with liquid water oceans, supporting its status as a target molecule for remotely searching for life elsewhere (e.g., Des Marais et al. 2002). Over the past decade, numerous studies have challenged this view by proposing potential mechanisms for the abiotic accumulation of atmospheric $O_2$ on exoplanets (e.g., reviews in Meadows 2017; Meadows et al. 2018). We summarize these mechanisms and the potential spectral features that could disambiguate potential "false positives" from true positive biosignatures. Because $O_2$ is the most well-studied potential remote biosignature, our treatment of its potential false positives accompanying spectral signatures is more comprehensive than we present for other potential biosignature gases.

Figure 5 provides a graphical summary of proposed $O_2$ false positives and the atmospheric context that could identify them. It is important to note that, as of this writing, the characterization of the atmospheric composition of terrestrial exoplanets is just beginning. Therefore, we have limited information about the prevalence of abundant abiotic $O_2$ on exoplanets, just as we are limited in understanding the prevalence of alien biospheres.

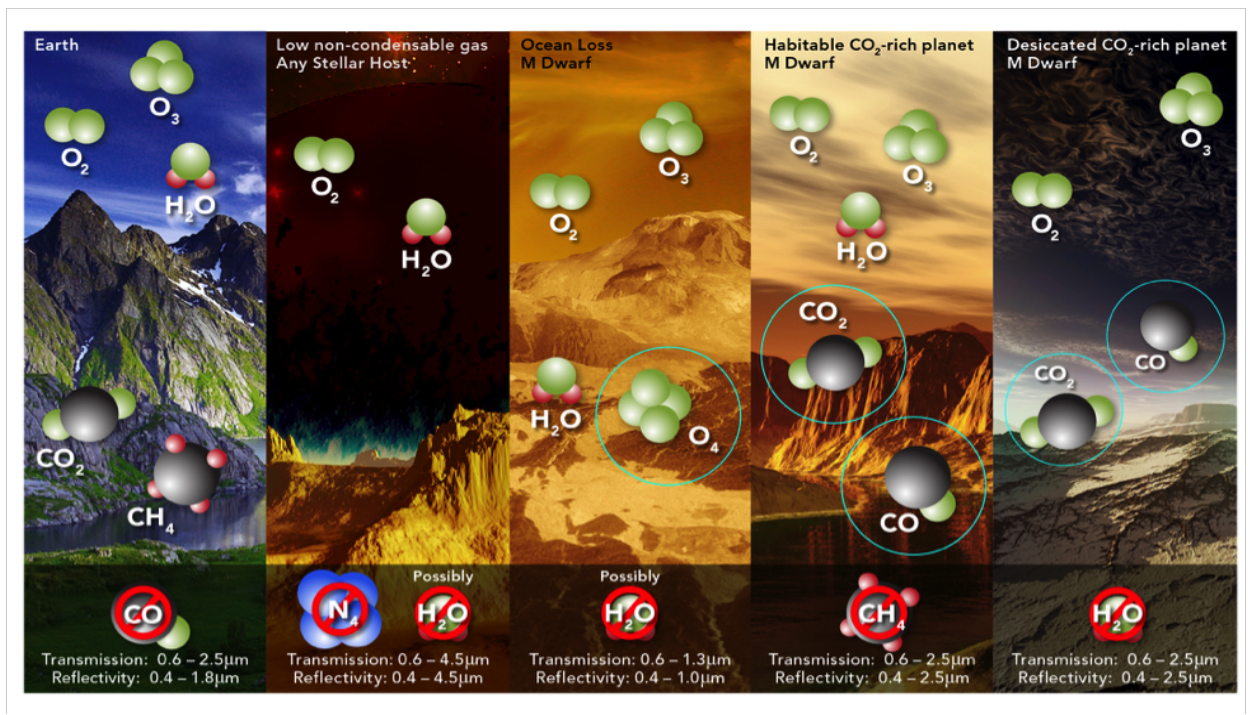

**Figure 5.** Graphical illustration of potential scenarios leading to potentially detectable abiotic $O_2$ and/or $O_3$. Circle molecules identify spectrally active species such as $O_4$ ($O_2$-$O_2$ collisionally induced absorption) that could be used to fingerprint the origin(s) of $O_2$ and/or $O_3$, while the red strike-through indicates species that should be absent. Modern Earth is distinguished by the combination of spectrally active $O_2$, $O_3$, $H_2O$, $CO_2$, and $CH_4$ together with low levels of CO. See section "False Positives for Oxygen and Ozone Biosignatures" in this text for additional details. This figure is reproduced from Meadows et al. (2018) under Creative Commons Attribution License CC-BY. Graphic artist: R. Hasler.

**_Photochemical $O_2$ and $O_3$ in $CO_2$-rich Atmospheres._** The photolysis of O-bearing molecules such as $CO_2$ can instantiate a series of photochemical reactions that generate $O_2$ and $O_3$ molecules. Indeed, the dry atmosphere of Mars has low but detectable (from Earth and interplanetary space) levels of $O_2$ and $O_3$ from oxygen liberated by $CO_2$ photolysis (Noxon et al. 1976; Fast et al. 2006). The direct recombination of CO and O into $CO_2$ (CO + O + M → $CO_2$ + M) is spin forbidden, so it occurs slowly. The photolysis products from abundant water vapor ($H_2O$ + hν → H + OH) can effectively catalyze this recombination, however:

$$CO + OH \rightarrow CO_2 + H$$
$$O_2 + H + M \rightarrow HO_2 + M$$
$$HO_2 + O \rightarrow O_2 + OH$$
$$\text{Net: } CO + O \rightarrow CO_2$$

The Martian atmosphere's water vapor prevents the decomposition of its $CO_2$ into CO and $O_2$ (McElroy and Donahue 1972; Krasnopolsky 2011). On Venus, the CO and O recombination is catalyzed by volcanic HCl (DeMore and Yung 1982). A very dry terrestrial exoplanet with a $CO_2$-rich atmosphere, extremely low $H_2O$ content, and the lack of other chemical catalysts could hypothetically decompose into a CO and $O_2$-rich atmosphere, depending on other factors (Gao et al. 2015). Some authors have found that the stellar spectrum of M dwarf host stars can drive the production of abiotic $O_2$ (and $O_3$) even on planets with surface oceans and abundant atmospheric water vapor (Tian et al. 2014; Domagal-Goldman et al. 2014; Harman et al. 2015). M dwarf host stars have more intense FUV radiation, which drives $CO_2$ photolysis, but less of the NUV radiation that is critical for photolyzing water and initiating the catalytic cycles that recombine CO and O in otherwise anoxic atmospheres. However, these results are sensitive to model assumptions, including the input disassociation cross-sections (Ranjan et al. 2020) and the lightning flash rate (Harman et al. 2018). $NO_x$ species (e.g., NO, $NO_2$) produced by lightning can drive the catalytic recombination of CO and O (Harman et al. 2018). The time-dependent UV radiation and particle fluxes from flares are also important, though currently understudied, considerations when modeling potential false positives for $O_2$ and $O_3$ (France et al. 2013, 2016; Chen et al. 2020).

Ranjan et al. (2023) have found that some past results predicting robust abiotic $O_2$ accumulation from $CO_2$ decomposition on terrestrial planets orbiting M dwarf host stars were erroneous and caused by instituting a model "top" at too low of an altitude (too high of a pressure), which effectively confined $CO_2$ photolysis to one numerical layer. They argue that abiotic $O_2$ accumulation via $CO_2$ photolysis will not generate $O_2$ levels above 1%, and perhaps much lower, assuming abundant tropospheric water vapor (i.e., habitable conditions). However, these $O_2$ levels can generate spectrally apparent $O_3$, which may be confounding when using $O_3$ as a tracer for biotic $O_2$. This false positive scenario could be identified via spectrally active CO (with features at 1.6, 2.35, and 4.7 μm), which accumulates before abiotic $O_2$ begins to rise (Schwieterman et al. 2016). Very dry atmospheres that facilitate abiotic $O_2$ accumulation via lack of $HO_x$ (e.g., OH, $HO_2$) catalysts could be identified via the lack of water vapor absorption (Gao et al. 2015). Future theoretical work—and, ultimately, observations—will further elucidate the extent of abiotic $O_2$ and $O_3$ that can be maintained in terrestrial atmospheres from the photolysis of other O-bearing species.

***Hydrogen Loss and Oxygen Retention During Early Runaway Greenhouses.*** The evolution of the atmospheres of terrestrial planets can be significantly influenced by the early, pre-main-sequence (PMS) phase of stellar evolution. During the PMS, protostars have yet to begin fusing hydrogen in their cores, and their luminosity is instead powered by gravitational contraction (Kippenhahn et al. 1990). As the radii of the stars shrink, their luminosity can decrease by one to two orders of magnitude before entering the main sequence, the core hydrogen-burning phase. The PMS phase for M dwarf stars ($0.08\ M_\odot \le M \le 0.6\ M_\odot$) can last hundreds of millions of years, compared to ~50 Myr for the Sun (Baraffe et al. 1998; Reid and Hawley 2005). Consequently, planets in the habitable zones of such stars today likely experienced extended runaway greenhouse states, which would have enriched their upper atmospheres with water vapor.

In addition to their extended PMS phases, due to their fully convective interiors and consequent magnetic activity, M dwarfs emit much more X-ray and extreme ultraviolet (XUV) radiation than Sun-like stars. This XUV radiation can contribute to substantial atmospheric loss, potentially ablating away entire atmospheres (Lammer et al. 2003, 2008; Johnstone et al. 2019) and—coupled with photodissociation of atmospheric water vapor—leading to the equivalent loss of many oceans of water (Ramirez and Kaltenegger 2014; Luger and Barnes 2015; Tian 2015). Because hydrogen atoms are less massive than oxygen atoms, hydrogen can more easily escape to space, though hydrodynamic flows can also carry away oxygen. Each Earth Ocean is equivalent to ~240 bars of $O_2$, so, depending on initial conditions, hundreds to thousands of bars of $O_2$ can be left behind to oxidize the planet's interior, surface, and atmosphere (Luger and Barnes 2015; Tian 2015).

Potential planetary sinks for this atmospheric oxygen (other than hydrodynamic loss) include dissolution in a magma ocean coupled with oxidation of reduced iron (Zahnle et al. 1988; Chassefière et al. 2012; Hamano et al. 2013), non-thermal ion escape (Persson et al. 2020), and reaction with reducing volcanic gases over time. Venus went through a runaway greenhouse at some point in its history, which could have removed more than an Earth ocean of water (Kasting et al. 1984; Kasting 1988; Chassefière 1996; Gillmann et al. 2009). If so, Venus would have at one time accumulated abundant abiotic atmospheric $O_2$, which was then consumed through some combination of the sinks enumerated above (Chassefière et al. 2012; Kane et al. 2019). Given the likely broad distribution in initial conditions among terrestrial planets, it was predicted that some Venus-like post-runaway atmospheres may have maintained elevated $O_2$ levels today (Chassefière et al. 2012).

Luger and Barnes (2015) demonstrated that planets currently within the habitable zones of M dwarf stars are particularly susceptible to abiotic $O_2$ accumulation during their star's pre-main sequence phase when the stars' luminosities and XUV fluxes are substantially enhanced relative to later times. Depending on their initial water endowments, these planets could become completely desiccated or retain substantial $H_2O$, complicating habitability and biosignature assessments. Many additional authors have predicted the possibility of high levels (tens, hundreds, or thousands of bars) of abiotic $O_2$ in the atmospheres of terrestrial planets orbiting M dwarf stars both within and outside the habitable zone (Schaefer et al. 2016; Wordsworth et al. 2018; Krissansen-Totton and Fortney 2022). Near-future characterization of "Venus zone" exoplanets (Kane et al. 2014) may elucidate the prevalence of post-runaway $O_2$-rich atmospheres in the solar neighborhood, which will have important implications for future biosignature surveys (Ostberg et al. 2023).

Abiotic $O_2$-rich post-runaway atmospheres could be identified by $O_2$ abundances that are too large to be biologically produced. On Earth, $O_2$ levels are self-regulated over time by negative feedbacks, including combustion of terrestrial vegetation (i.e., fires) to levels $\lesssim 0.35$ bar (Lenton 2013). High levels of $O_2$ could be fingerprinted by $O_2$-$O_2$ collisionally induced absorption (CIA), sometimes called $O_4$, which is strongly dependent on density (Misra et al. 2014; Schwieterman et al. 2016). $O_2$-$O_2$ CIA has prominent absorption features at 0.345, 0.36, 0.38, 0.445, 0.475, 0.53, 0.57, 0.63, 1.06, 1.27, and 6.4 μm (Hermans et al. 1999; Maté et al. 2000; Richard et al. 2012; Karman et al. 2019). At optical and near-infrared wavelengths, these $O_2$-$O_2$ bands would generate large absorption features in reflected light (Figure 6; Meadows et al. 2018). In transit, the 1.06 and 1.27 μm bands would be the best target for identifying $O_2$-$O_2$ CIA (Lustig-Yaeger et al. 2019). The 6.4 μm band could also fingerprint high-$O_2$ atmospheres in transit, assuming minimal interference from water vapor absorption (Fauchez et al. 2020).

***Hydrogen Loss and Oxygen Retention in Atmospheres with Limited Non-Condensable Gases.*** Abiotic oxygen could accumulate in a terrestrial planet's atmosphere without going through a runaway greenhouse. On Earth, almost all water vapor is confined to the troposphere via the "cold trap" at the tropopause, which is the altitude of the temperature minimum where the troposphere transitions into the stratosphere. The tropopause altitude depends on latitude and ranges from 9 km at the poles to 17 km at the equator. Water condenses at the tropopause's temperature and pressure conditions, resulting in a relatively dry stratosphere. However, the efficacy of the cold trap relies on the presence of non-condensing gases like $N_2$ and argon (Ar). Wordsworth and Pierrehumbert (2014) show that reducing the non-condensing gas abundances below ~0.2 bar lifts the cold trap and leads to the accumulation of water vapor in the stratosphere, where UV photons can more readily photolyze $H_2O$. The hydrogen liberated from this $H_2O$ photolysis can diffuse through the upper atmosphere and escape to space, resulting in the net oxidation of the planetary surface and/or atmosphere. Once $O_2$ accumulates such that the total pressure of non-condensing gases is $\gtrsim 0.2$ bar, the cold trap is (re-)established, limiting further H loss and $O_2$ build-up. (The rate at which $O_2$ builds up in the atmosphere depends on additional factors, including crustal oxidation and the outgassing of volcanic gases.) This amount of $O_2$ is practically identical to Earth's total $O_2$ abundance of 0.21 bar.

This abiotic $O_2$ scenario could be positively identified or ruled out by constraining the abundance of non-condensing gases. $N_2$ is difficult to detect because it is a homonuclear molecule with no transitional dipole moment; however, $N_2$ could be directly identified via its $N_2$-$N_2$ CIA band centered at 4.3 μm (Lafferty et al. 1996; Schwieterman et al. 2015b), which produces potentially detectable signatures in Earth's transmission spectrum (Lustig-Yaeger et al. 2023) and could plausibly be detected in the transmission spectra of exoplanets (Kaltenegger et al. 2020). Directly detecting $N_2$ would be difficult in reflected light because of the absence of significant absorption features. However, $N_2$ could be inferred indirectly in reflected light by characterizing the total atmospheric pressure via Rayleigh scattering and/or pressure broadening and ruling out other absorbing gases through a process of elimination, given sufficient spectral coverage (Hall et al. 2023).

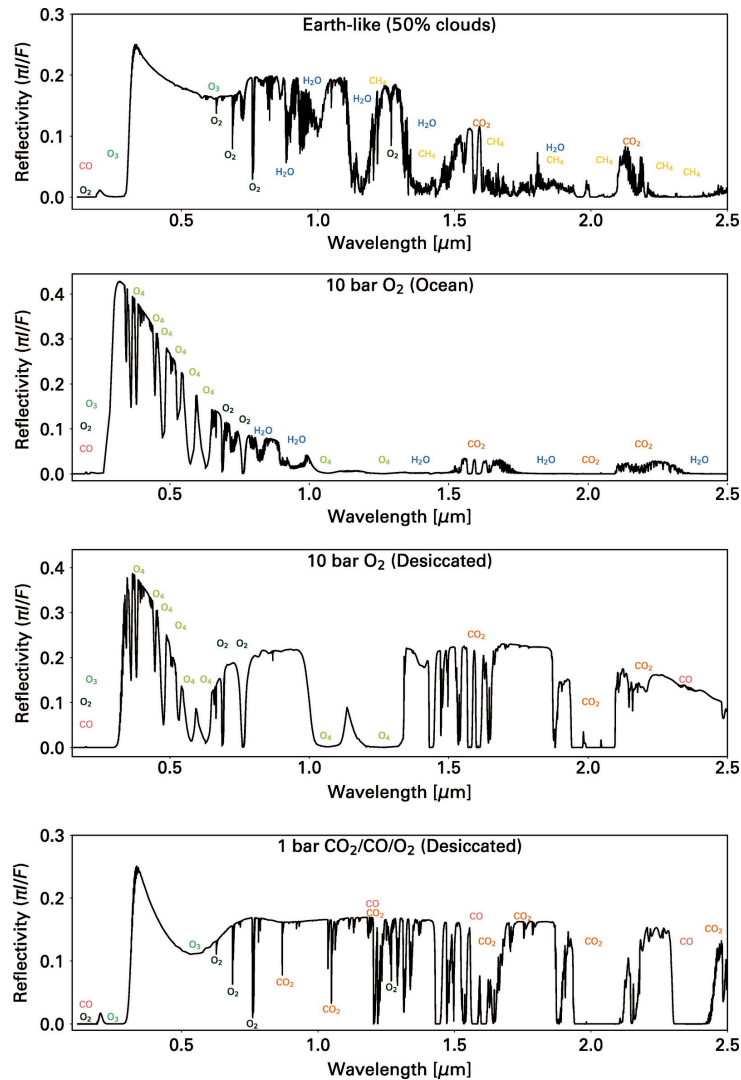

**Figure 6**. Spectra of hypothetical atmospheres for Proxima Centauri b in reflected light, illustrating spectral features of oxygen consistent with an inhabited planet (topmost panel) or various false positive mechanisms (all other panels). **First panel:** photochemically self-consistent Earth, with enhanced $CH_4$ spectral features consistent with longer photochemical lifetimes predicted for planets orbiting M dwarf host stars. **Second panel:** 10 bar abiotic $O_2$-dominated atmosphere with a remaining surface ocean. **Third panel:** 10 bar abiotic $O_2$-dominated atmosphere after complete desiccation and loss of surface ocean. **Final panel:** desiccated, $CO_2$-rich atmosphere photochemically decomposed into $CO_2/CO/O_2$ due to lack of catalysts requiring H. Compare with conceptual illustrations in Figure 5. This figure is sourced from Meadows et al. (2018a, b) under Creative Commons Attribution License CC-BY.

***Perpetual Greenhouses, Water Worlds, and Desert Worlds.*** Krissansen-Totton et al. (2021) proposed three other scenarios for the abiotic accumulation of $O_2$ on terrestrial worlds orbiting sun-like stars, all of which relate to the initial volatile inventories of the planets in question. They find that if the initial $CO_2$:$H_2O$ inventory exceeds one by mass, the planet's atmosphere becomes trapped in a perpetual greenhouse state that does not abate over several Gys for worlds at 1 AU, a scenario also proposed by prior authors (Marcq et al. 2017; Salvador et al. 2017). Because the atmospheric water is supercritical, silicate weathering reactions requiring liquid water that would otherwise draw down $CO_2$ cannot operate effectively. As in the runaway greenhouse scenarios described above, atmospheric water vapor can be photolyzed, followed by hydrogen escape and atmospheric oxidation. The lack of liquid water further arrests oxidative weathering that would consume this abiotic $O_2$ in the atmosphere, potentially allowing it to accumulate to high concentrations (i.e., $p_{O2} > 0.02$ bar). However, this scenario requires a low crustal oxidation efficiency, as atmospheric $O_2$ could directly oxidize fresh crust (Krissansen-Totton et al. 2021). Strong $CO_2$ bands and lack of a habitable surface could identify this scenario. Because the structure of absorption bands is temperature-dependent, their shape could also indicate temperatures too high for habitability (Young et al. 2024).

A water world scenario, assuming initial $H_2O$ inventories of over 10 Earth oceans, may also lead to the accumulation of abiotic $O_2$ (Krissansen-Totton et al. 2021). This is caused by the pressure overburden of a massive ocean arresting the creation of new crust, which strongly limits surface oxygen sinks (Noack et al. 2016; Kite and Ford 2018). In this case, $O_2$ is net produced slowly due to hydrogen escape but accumulates over geologic time due to strongly reduced sinks. This false positive scenario could be indicated by the absence of continents, which could be revealed via rotational mapping techniques (Cowan et al. 2009; Kawahara and Fujii 2011; Fujii et al. 2017; Lustig-Yaeger et al. 2018).

Finally, the last scenario presented by Krissansen-Totton et al. (2021) is a desert world scenario, predicated on very low initial volatile inventories, which we briefly summarize as follows. In this case, the low levels of volatiles generate a weak early greenhouse, allowing the underlying magma ocean to cool in less than ~$10^5$ years. Consequently, abiotically generated atmospheric $O_2$ sourced from XUV-driven hydrogen escape cannot react with the reductants in the magma ocean. This $O_2$ could persist over Gyrs because the low volatile inventory implies a small source of volcanic reductants, but this scenario also requires inefficient crustal oxidation. This desert world scenario could be identified by the lack of ocean glint (Gaidos and Williams 2004; Williams and Gaidos 2008; Zugger et al. 2010; Robinson et al. 2014; Lustig-Yaeger et al. 2018; Ryan and Robinson 2022).

Importantly, for each scenario listed above, the extent of $O_2$ accumulation depends on many factors that can differ drastically for exoplanets, and many combinations of these variables do not lead to predicted abiotic $O_2$ accumulation. Moreover, each scenario is accompanied by predicted diagnostic features that could either identify or rule out the scenario in question. Finally, in every case, the co-existence of $O_2$ with complementary reduced biosignature gases, such as $CH_4$ and/or $N_2O$, would be a distinguishing factor that would exclude an abiotic origin for $O_2$.

## Methane ($CH_4$)

Methanogenesis is an ancient metabolism that arose early in the history of life on Earth and likely contributed to an early methane-rich atmosphere in the Archean Eon (Ueno et al. 2006;

Catling and Zahnle 2020; Stüeken et al. 2024). Methanogenesis involves the chemotrophic generation of energy via anaerobic respiration that results in the production of methane gas. Most commonly, $CO_2$ is used as a terminal electron acceptor, or acetic acid ($CH_3COOH$) is disproportionated into $CH_4$ and $CO_2$. These pathways can be simplified as:

$$CO_2 + 4H_2 \rightarrow CH_4 + 2H_2O$$
$$CH_3COOH \rightarrow CH_4 + CO_2$$

where $H_2$ is hydrogen gas. Methanogenesis also requires catalysis by specialized enzymes in complex biochemical processes not captured by these simplified net equations (Ferry 1999).

Methanogenesis occurs in various anaerobic environments today, including in oxygen-poor soils, in anoxic lakes, at hydrothermal vents, and in the gastrointestinal tracts of animals (Thauer et al. 2008; Martin et al. 2008). Importantly, $CH_4$ can be oxidized aerobically and anaerobically by other organisms, such as *Methylococcus* spp. (Hanson and Hanson 1996) and sulfate reducers (Cui et al. 2015), respectively, before it is released into the atmosphere. Therefore, the total biogenic production of $CH_4$ on Earth is greater than its net atmospheric release. This net flux on Earth is approximately ~30-40 Tmol/year (Dlugokencky et al. 2011), inclusive of biotic, industrial, and abiotic sources. This results in a modern $CH_4$ concentration of ~1.9 ppm (Lan et al. 2023), though pre-industrial concentrations were likely closer to 500 ppb (Solomon et al. 2007). Sources of abiotic methane on Earth include water-rock reactions such as serpentinization, which could account for as much as 10% of the non-anthropogenic methane production (Etiope and Sherwood-Lollar 2013).

$CH_4$ is a biosignature in Earth's atmosphere because it has strong thermodynamic and kinetic disequilibrium with atmospheric $O_2$ (Lovelock 1965; Hitchcock and Lovelock 1967; Sagan et al. 1993). The atmospheric lifetime of methane on Earth is about 10 years. Its major photochemical sink is the hydroxyl radical (OH), which in Earth's $O_2$-rich atmosphere is primarily sourced from the photolysis of tropospheric ozone ($O_3 + h\nu \rightarrow O(^1D) + O_2$ followed by $O(^1D) + H_2O \rightarrow 2OH$; Jacob 1999). The atmospheric lifetime of $CH_4$ is a strong function of the incident stellar spectrum, especially in the context of $O_2$-rich atmospheres, and the same $CH_4$ flux could result in $CH_4$ concentrations ~1,000 times greater for Earth-like terrestrial planets orbiting M dwarf stars (Segura et al. 2005; Rugheimer et al. 2015; Wunderlich et al. 2019). This is because M dwarfs emit substantially less energy at NUV wavelengths that photolyze ozone compared to sunlike stars, corresponding to diminished production of OH radicals to react with and remove $CH_4$ (Segura et al. 2005).

Methane has strong absorption features in the NIR at 1.65, 2.3, and 3.3 μm (with weaker bands at 0.9, 1, 1.15, and 1.35 μm) and in the MIR most strongly in a wide band centered at ~7.7 μm and extending from ~7.2-8.4 μm. These bands often overlap strongly with $H_2O$, so moderate to high spectral resolution is required to distinguish these species. We can search for $CH_4$ in terrestrial planets' reflection, transmission, or emission spectra, with NIR bands more favorable for reflected or transmitted light observations and MIR bands most favorable for emitted light observations. Importantly, Earth's modern concentration of methane may be undetectable in reflected light, because of its low concentration and correspondingly weak absorption bands; however, this was not the case for the Archean or perhaps the Proterozoic Eons, with higher methane concentrations (Reinhard et al. 2017; The LUVOIR Team 2019; Gaudi et al. 2020). M and K dwarf host stars, with their corresponding stellar spectra and consequent implications for photochemistry, may promote detectable concentrations of $CH_4$ in modern Earth-like atmospheres in reflected light (e.g., Segura et al. 2005; Arney 2019).

Of course, plentiful abiotic methane also exists in the atmospheres of the solar system's gas planets, where it is made by equilibrium processes in high-temperature layers of $H_2$-rich atmospheres and transported to higher altitudes with lower temperatures (Lodders and Fegley 2002; Madhusudhan et al. 2016). On a temperate planet, such equilibrium reactions cannot happen in the atmosphere on a geologically relevant timescale, so an active surface source is necessary to maintain atmospheric methane. In an anoxic terrestrial atmosphere like the Archean Earth, $CH_4$ is destroyed by OH radicals sourced from water vapor photolysis or by direct photolysis of $CH_4$ by FUV radiation (Kasting et al. 1983). $CH_4$ is also prevalent in the atmosphere of Saturn's moon Titan, and while the origin of this $CH_4$ is debated (e.g., Glein 2015), its photochemical stability is aided by the cold temperatures (~90 K), which preclude abundant water vapor and downstream photochemical sinks like OH.

Krissansen-Totton et al. (2018b) suggest that the disequilibrium between $CH_4$ and $CO_2$ would constitute a biosignature if $CH_4$ mixing ratios exceed $10^{-3}$ to $10^{-2}$ and if both gases are observed without abundant CO. This is because $CO_2$ and $CH_4$ are two vastly different oxidation states of carbon. Any abiotic source (e.g., a mantle with oxygen fugacity orders of magnitude lower than modern Earth) that produces both should also generate abundant CO, which has an intermediate oxidation state. Indeed, the $CO_2$-$CH_4$ disequilibrium biosignature may have been observable in Earth's atmosphere for longer than observable $CH_4$-$O_2$ disequilibrium (Krissansen-Totton et al. 2018b). Coupled with the kinetic disequilibrium generated from the photochemical destruction of $CH_4$ in temperate atmospheres and the need for continuous sources, these factors motivate the potential of $CH_4$ to be an exoplanet biosignature even within atmospheres lacking observable $O_2$ (Thompson et al. 2022). It is important to note that in some scenarios, volcanic production could result in both $CO_2$ and $CH_4$ production, though CO would also be produced (Schaefer and Fegley 2010; Liggins et al. 2022, 2023). In addition, photochemical processes can produce $CO_2$ even in $H_2$-rich atmospheres that contain both $CH_4$ and $H_2O$. In this scenario, the carbon in $CH_4$ is photo-oxidized using the O liberated from $H_2O$ photolysis (Hu 2021); however, yet again, abundant CO would be predicted in this scenario as an intermediate product. It is also often assumed that CO would be drawn down by life via acetogenesis, as CO affords both a reductant and carbon source (Ragsdale 2004; Wang et al. 2016; Catling et al. 2018). Therefore, the observable absence (or low abundance) of CO underpins the potential utility of the $CO_2$-$CH_4$ biosignature pair in anoxic atmospheres. This factor, too, is complicated by the observation that life produces CO through various direct and indirect processes and does not need to consume CO at the maximum biotic rate (Schwieterman et al. 2019; Zhan et al. 2022).

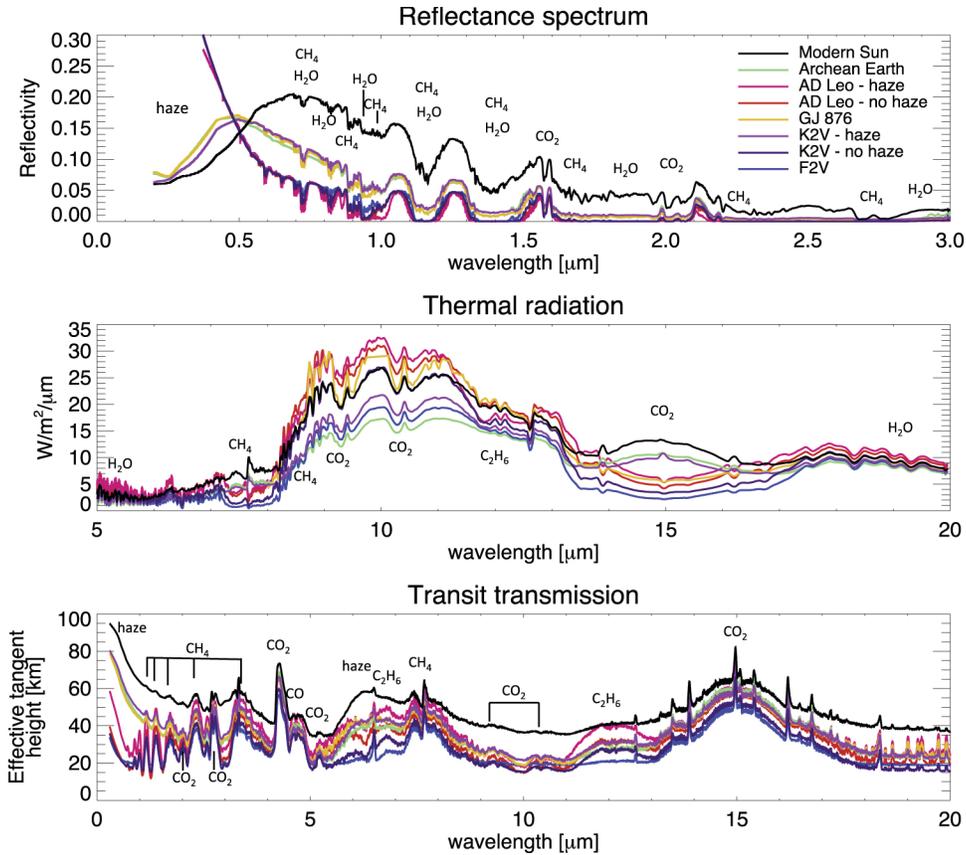

**Figure 7.** Synthetic spectra of anoxic Archean-like exoplanets orbiting other stars modeled with self-consistent photochemistry and climate. The top panel contains reflectance spectra, the middle panel contains thermal emission spectra, and the bottom panel contains transmission spectra. The K2V-haze case has a $CH_4/CO_2$ ratio 0.3, the AD Leo-haze case has a $CH_4/CO_2$ ratio of 0.9, and all other cases have $CH_4/CO_2$ ratios of 0.2. Used by permission of AAS, from Arney et al. (2017), *The Astrophysical* Journal, Vol. 836, Fig. 6, p. 49.

High levels of methane can lead to the production of organic hazes, which may have existed during portions of the Archean (Trainer et al. 2006; Zerkle et al. 2012; Arney et al. 2016) and may have been important for generating prebiotic molecules for Hadean Earth (Pearce et al. 2024). These hazes would have had both strong climatic and spectral impacts on Earth's atmosphere and could therefore be common on temperate exoplanets in a similar biogeochemical regime to Earth's Archean (Arney et al. 2017) or Hadean eons. For hydrocarbon haze to accumulate, the ratio of $CH_4$ to $CO_2$ must exceed ~0.1 (the exact ratio depends on the inherent photochemistry of the planet-star system), so it has been suggested that hydrocarbon haze could be an indirect exoplanet biosignature (Arney et al. 2018) given that it may be implausible for abiotic $CH_4$ production rates alone to generate haze in atmospheres with Archean-like levels of $CO_2$. The presence of haze at an unexpectedly low $CH_4/CO_2$ ratio could imply the presence of other organic molecules fueling haze formation (e.g. organic sulfur gases), even if these gases are not themselves detectable in the spectrum, which could strengthen a biological interpretation (Arney et al. 2018). However, organic haze is predicted for a wide variety of planetary scenarios, including those with no biological inputs (He et al. 2020; Moran et al. 2020; Gao et al. 2021). Nonetheless, photochemical models continue to predict that the distribution of carbon species ($CO_2$, $CO$, $CH_4$) will differ between abiotic and scenarios with a robust biosphere (Akahori et al. 2023; Watanabe and Ozaki 2024), though there may be substantial overlap, particularly for temperate anoxic planets orbiting M dwarf hosts. Figure 7 shows planetary spectra of hypothetical Archean-like planets orbiting various stars, displaying the simultaneous presence of $CH_4$, $CO_2$, organic haze, and ethane ($C_2H_6$) that results from the photochemical processing of $CH_4$ and organosulfur gases.

Though the interpretation of $CH_4$ biosignatures is complex—particularly in the absence of $O_2$ or $O_3$—it is nonetheless a key target in the search for life beyond Earth. Indeed, the $CO_2$-$CH_4$ biosignature couple may be the most observable potential biosignature in the TRAPPIST-1 planetary system (Krissansen-Totton et al. 2018a; Mikal-Evans 2021; Meadows et al. 2023).

## Nitrous Oxide ($N_2O$)

$N_2O$ is overwhelmingly produced by life on Earth due to microbial nitrogen metabolism in terrestrial and aquatic environments. Through the processes of nitrogen fixation, nitrification, and denitrification, atmospheric $N_2$ gas is converted into bioavailable forms of nitrogen such as ammonium ($NH_4^+$), oxidized to form nitrite ($NO_2^-$) and nitrate ($NO_3^-$), and then reduced back to $N_2$ gas, with $N_2O$ as an intermediate product (Thamdrup 2012; Tian et al. 2015). $N_2O$ is also produced through direct ammonia oxidation by some microorganisms (Prosser and Nicol 2012).

The modern concentration of atmospheric $N_2O$ is ~330 ppb, with a preindustrial concentration of ~270 ppb (Myhre et al. 2013). The primary atmospheric sink for $N_2O$ is photolysis, which converts $N_2O$ into $N_2$ and $O_2$, though reaction with atomic oxygen radicals also contributes to the photochemical destruction of $N_2O$. The photochemical survival of $N_2O$ is dependent on the incident stellar spectrum and oxygen abundance, as overlying $O_2$ (and accompanying $O_3$) can provide a shielding effect (Grenfell et al. 2014; Rugheimer and Kaltenegger 2018). The photochemical lifetime of $N_2O$ in Earth's atmosphere today is about 120 years (Prather et al. 2015). The net production of $N_2O$ by terrestrial and marine sourced on Earth today is ~0.4 Tmol/year (Tian et al. 2020). Figures 8 and 9 illustrate how $N_2O$ concentrations depend on molecular flux, $O_2$ concentration, and the incident stellar spectrum and provide a general example of how these factors can influence biosignature gas abundances.

The $N_2O$ concentration on Earth may have changed drastically over geologic time. The biogenic flux of $N_2O$ to the atmosphere depends on biological and environmental factors. The final step of the nitrogen cycle that converts $N_2O$ into $N_2$ gas is mediated by an enzyme called nitrous oxide reductase (Pauleta et al. 2013), which requires a copper catalyst. It has been suggested that $H_2S$-rich oceans in the Proterozoic Eon (2.5 to 0.5 Ga) would have been deficient in copper, reducing the efficiency of the last step in the denitrification cycle that converts $N_2O$ to $N_2$, which could have had substantial climatic implications given $N_2O$'s status as a greenhouse gas (Buick 2007; Roberson et al. 2011; Stüeken et al. 2024). In the absence of metal co-factors or the evolution of the nitrous oxide reductase enzyme (or an analog), the end member $N_2O$ emission flux would approach the total denitrification flux of the biosphere, which is on the order of ~20 Tmol $N_2$ per year (Canfield et al. 2010). However, nitrification and denitrification depend on various factors, including nutrient availability and oxygen concentrations so that this value could be smaller or larger on an exoplanet (Schwieterman et al. 2022).

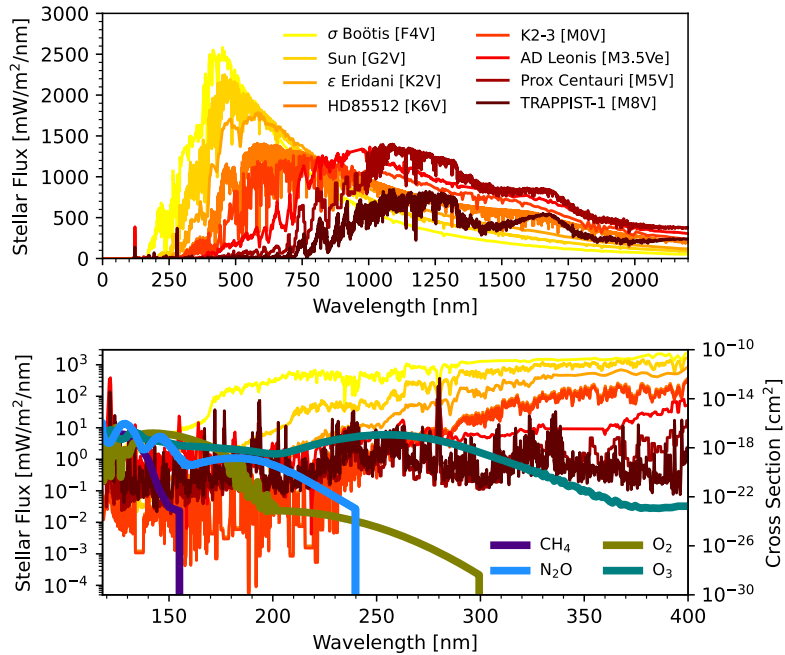

**Figure 8.** Top panel: Spectral flux received at an Earth-like insolation from various stellar host types in the UV-Vis-NIR. Bottom panel: the identical stellar spectra focused on the Far-UV to near-UV with the dissociation cross-section of $O_2$, $O_3$, $N_2O$, and $CH_4$ overplotted (right y-axis). The different energy distributions from each star will impact the photolysis rate and, thus, the steady-state concentration given a specific molecular production rate (see Figure 9). This figure was taken from Schwieterman et al. (2022) under Creative Commons Attribution License CC-BY.

$N_2O$ is advantageous as a potential remote biosignature due to its overwhelming biological origin on Earth and distinct spectral features in the near- and mid-IR (Sagan et al. 1993; Kaltenegger and Selsis 2007; Rauer et al. 2011). $N_2O$ can be seen weakly in Earth's near-infrared reflected light spectrum (Fig. 2; Sagan et al. 1993), and more strongly in Earth's emitted mid-infrared spectrum (Fig. 2; Mettler et al. 2023), and in Earth's transmitted light spectrum (Fig. 3; Lustig-Yaeger et al. 2023). $N_2O$ has significant absorption bands in the near-IR at 2.25, 2.9, 4.0, and 4.5 μm, with weaker bands at 1.5, 1.6, 1.7, 1.8, 2.6, and 3.7 μm (yet weaker bands are present at shorter wavelengths). In the mid-IR, $N_2O$'s strongest bands reside at 7.8, 8.5, and 17 μm, and would be compelling targets for space-based direct imaging of terrestrial worlds (Figure 10; Angerhausen et al. 2024). Importantly, many of these absorption bands overlap with $CH_4$; for example, the 7.8 μm $N_2O$ band overlaps with the 7.7 μm $CH_4$ band, so the actual detectability of the gas will depend on a confluence of factors, including the concentrations of overlapping absorbing species.

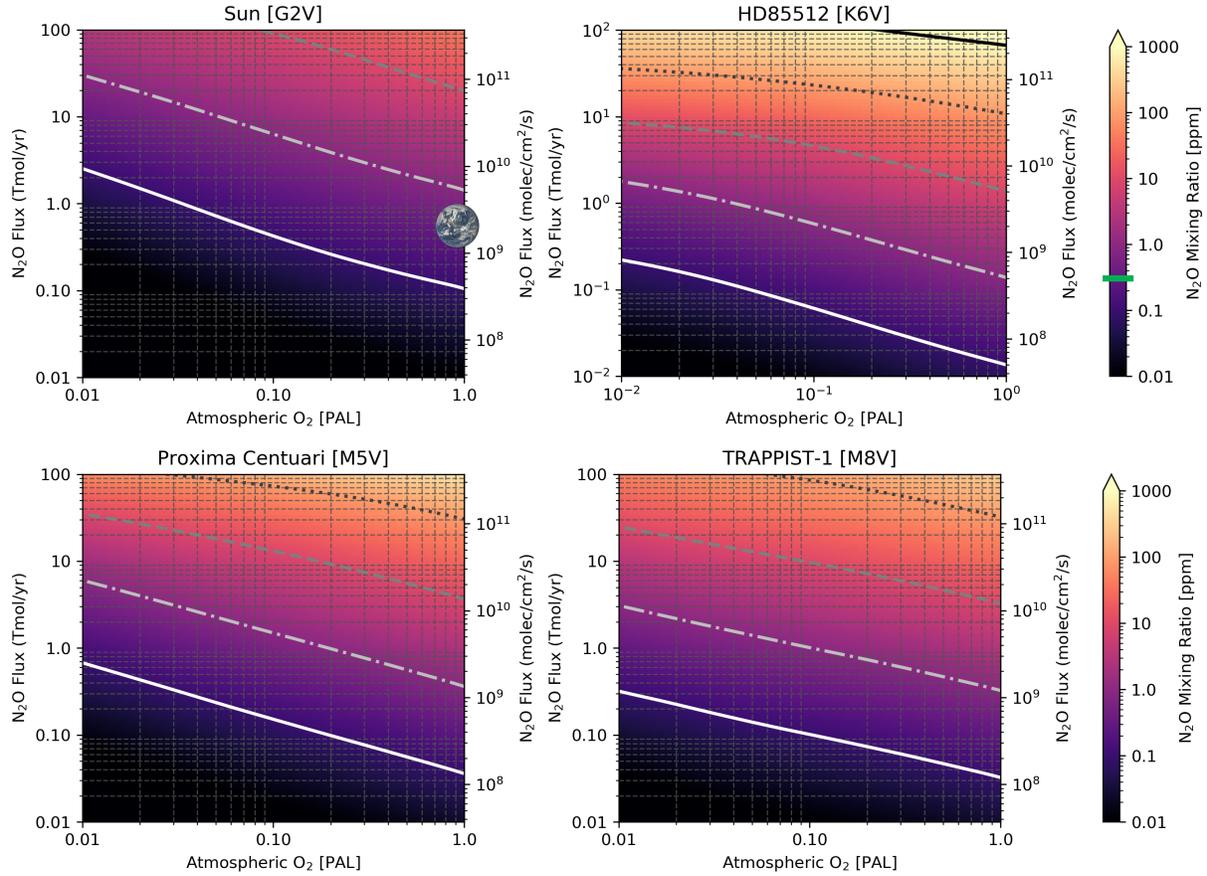

**Figure 9.** $N_2O$ concentrations in parts-per-million (ppm) as a function of $N_2O$ flux (y-axes) and atmospheric $O_2$ (x-axes) predicted by a photochemical model for an Earth-like planet orbiting within the habitable zone of each named star. The stars listed here correspond to those whose spectra are shown in Figure 8. The inset Earth image in the **top left panel** indicates Earth's $N_2O$ molecular flux (0.4 Tmol/yr; $1.5 \times 10^9$ molecules/cm$^2$/s). The green horizontal line indicates Earth's current $N_2O$ atmospheric mixing ratio (~330 parts-per-billion; $3.3 \times 10^{-7}$). Data subpanels were adapted from Schwieterman et al. (2022) under Creative Commons Attribution License CC-BY.

Figure 10 shows the resulting planetary spectra of Earth-like planets with varied $N_2O$ production rates and stellar hosts. These scenarios correspond to the ones shown in Figures 8 and 9, and together, this sequence of three figures traces the vertically integrated connection between the intrinsic properties of the $N_2O$ molecule (i.e., dissociation cross-sections), incident stellar spectra, biosignature molecular fluxes, and resulting spectral observables. These spectra also illustrate that $N_2O$ may be a more detectable biosignature on planets orbiting cooler stars than the Sun. A hypothetical inhabited TRAPPIST-1e with an $N_2O$ flux constituting a significant fraction of Earth's planetary denitrification flux (see Stüeken et al. 2024) would produce $N_2O$ transit signatures equivalent to other major gas absorbers (e.g., $CO_2$, $CH_4$), and could be detectable by JWST (Schwieterman et al. 2022).

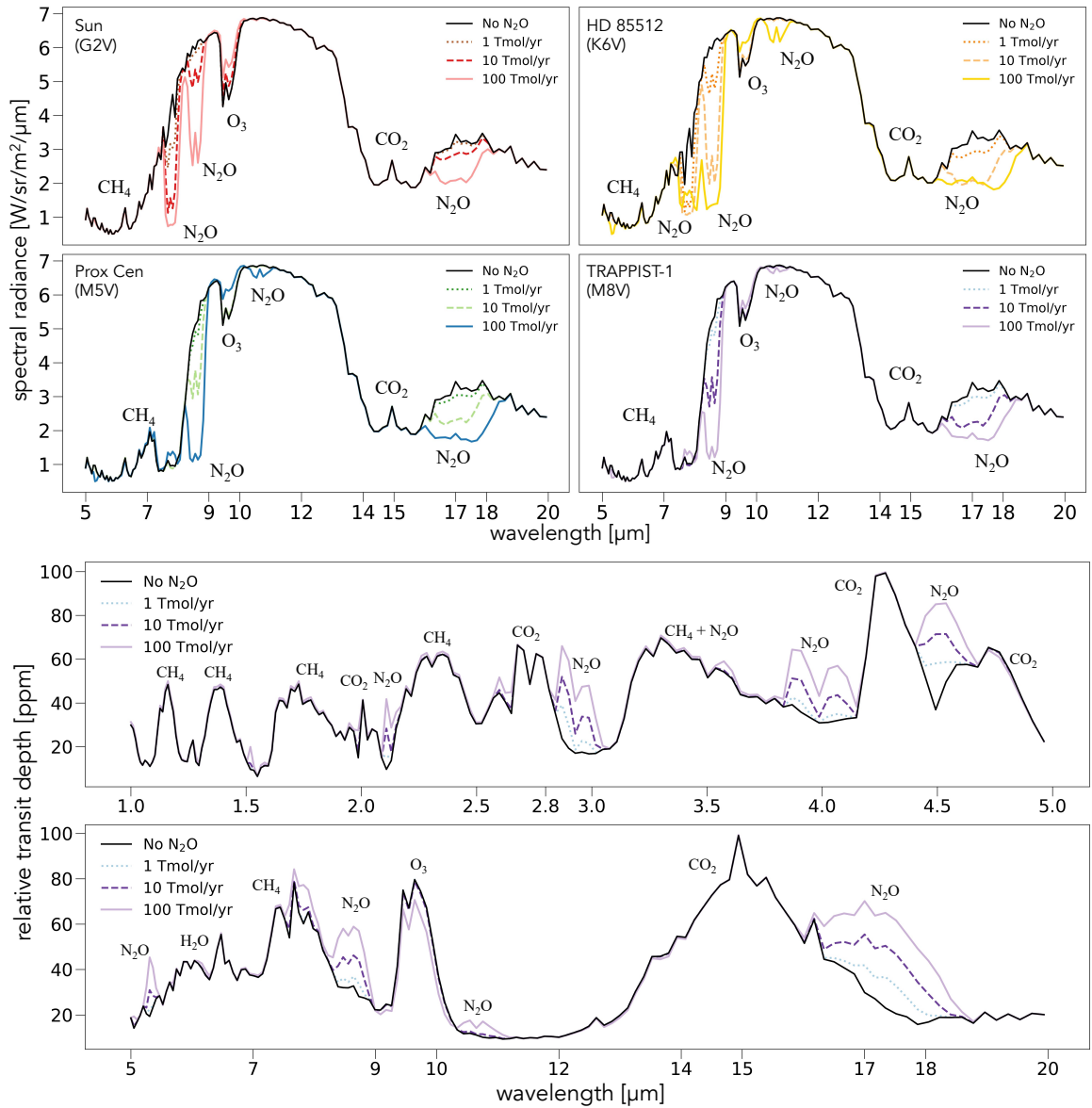

**Figure 10.** Spectra of Earth-like planets showing $N_2O$ features as a function of $N_2O$ production rate. The top set of four panels shows the IR emission spectra of Earth-like planets with 21% $O_2$, 50% cloud cover, and net $N_2O$ fluxes of 1, 10, and 100 Tmol/yr. The bottom two panels show a hypothetical transmission spectrum of TRAPPIST-1e under the same assumptions. Major absorbing species are indicated. Each scenario's abundance is determined from the photochemical simulations shown in Figure 9. These figures were adapted from Schwieterman et al. (2022) under <u>Creative Commons Attribution License CC-BY</u>.

The abiotic sources of $N_2O$ are limited on Earth, with a very small amount produced by lightning  (Schumann and Huntrieser 2007) and chemodenitrification (Samarkin et al. 2010; Jones et al. 2015) processes. Chemodenitrification refers to the abiotic reduction of oxidized aqueous nitrogen species (e.g., $NO_3^-$) by reductants such as ferrous iron ($Fe^{2+}$), which can produce $N_2O$ gas abiotically (Stanton et al. 2018). However, $N_2O$ production via chemodenitrification requires the simultaneous presence of reductants and oxidants in ocean layers. This is implausible with an abiotic $O_2$ atmosphere, which requires a completely oxidized ocean for long-term stability. Lightning can produce small amounts of $NO_x$ species that can be chemodenitrified without biotic $O_2$. However, while chemodenitrification could have maintained $N_2O$ concentrations in the ~10 ppb range in the Archean (Buessecker et al. 2022), remotely detectable concentrations likely require direct or indirect biological production (Schwieterman et al. 2022).

Photochemical simulations show that an active star can produce abundant StEPs that could split the $N_2$ molecule and initiate a series of photochemical reactions that could produce abiotic $N_2O$ (Airapetian et al. 2016, 2020; Chen et al. 2020). However, such a scenario could be identified by the characterization of the star and the additional spectrally active gaseous species that would be produced, such as $NO_2$, $HNO_3$, and HCN. The simultaneous presence of $N_2O$ and $O_2$ would be a stronger biosignature than either alone, akin to the $O_2$ and $CH_4$ biosignature couple.

Schwieterman et al. (2022) present a more comprehensive summary of the potential for $N_2O$ biosignature detection on exoplanets, including the plausible range of biological fluxes, photochemical accumulation, and resulting spectral signatures in transmission and emission spectra. This work also includes a more extensive discussion of possible abiotic sources for $N_2O$ and the possible distinguishing spectral features of those scenarios.

## Alternative biosignature gases

Above, we have reviewed the potential atmospheric biosignatures that exist at spectrally significant abundances ($\gtrsim$ 100 ppb) in the modern atmosphere and likely generated remotely detectable signatures for a substantial fraction of Earth's geologic history ($\gtrsim$ 500 Myr; also see Table 1). Life produces more volatile compounds that do not accumulate to greater concentrations in Earth's atmosphere due to lower net biological production and/or shorter photochemical lifetimes. However, the net production of biogenic gases can be highly sensitive to the chemical nature of the local environment and evolutionary contingencies (e.g., Garcia et al. 2020; Parsons et al. 2021), and the photochemical lifetime is a strong function of the host star (e.g., Segura et al. 2005; France et al. 2016). Given the likely diverse compositions, evolutionary histories, and astrophysical settings of rocky exoplanets, biosignature gases at trace concentrations on Earth could have significant abundances elsewhere. Moreover, alternative biochemistries (i.e., "life as we don't know it") may further expand the range of potential biosignatures (Davila and McKay 2014; Seager et al. 2016; Chan et al. 2019; Petkowski et al. 2020). Below, we describe proposed alternative biosignature gases (Table 2) and provide the most recent references for comprehensive evaluations of their biological production, sources of abiotic false positives, and potential for atmospheric accumulation and detectability on exoplanets.

**Table 1.** Abundant and/or Prominent Earth Atmospheric Biosignature Molecules

| Biosignature | Production Environment | Concentration on Earth | False Positives and Secondary Outcomes | Main Spectral Features (μm) (strongest **bolded**) Via HITRAN and NIST | Citation (see references therein) |
|---|---|---|---|---|---|
| Oxygen ($O_2$) | Photic terrestrial and marine environments, produced by cyanobacteria, algae, and plants | 20.95% | • Low noncondensable inventories (Wordsworth and Pierrehumbert 2014)<br>• Ocean Loss (Luger and Barnes 2015; Tian 2015)<br>• $CO_2$ photolysis (Domagal-Goldman et al. 2014; Gao et al. 2015; Harman et al. 2015)<br>• High $CO_2/H_2O$ inventories or extremely low $H_2O$ inventories (Krissansen-Totton et al. 2021)<br>• Atmospheric Exchange/Exogenous Delivery (Felton et al. 2022) | $O_2$ at 0.63, 0.69, **0.76**, 1.27 CIA/$O_4$ at 0.445, 0.475, 0.53, 0.57, 0.63, 1.06, **1.27**, **6.4** | Meadows 2017; Meadows et al. 2018b |
| Ozone ($O_3$) | Photochemical product of $O_2$, $H_2O$, $SO_2$, $CO_2$ | 0.07 ppm | • Abiotic $O_2$ or $CO_2$ can produce $O_3$ features (Domagal-Goldman et al. 2014; Gao et al. 2015; Harman et al. 2015) | 0.25 (Hartley), 0.4-0.7 (Chappuis), 2.6, 4.7, **9.65**, 14.6 | Meadows et al. 2018b; Kozakis et al. 2022 |
| Methane ($CH_4$) | Wetlands, rice paddies, livestock emissions, biomass burning & decomposition | 1.90 ppm | • Serpentinization<br>• Fischer-Troph reactions<br>• high-temperature mantle reactions<br>• equilibrium reactions in high-T regions of gas giants | 1.65, 2.4, 3.3, **7.7** (broad) | Etiope and Sherwood-Lollar 2013; Thompson et al. 2022 |
| Nitrous Oxide ($N_2O$) | Marine settings, some soils | 0.33 ppm | • Production by StEPs<br>• Chemodenitrification<br>• trace production by lightning and $NO_x$ reactions | 1.5, 1.6, 1.7, 1.8, 2.3, 2.6, 2.9, 3.7, **4.0**, **4.5**, **7.8**, **8.5**, 17 | Schwieterman et al. 2022 |

| **Table 2.** Alternative Atmospheric Biosignatures | | | | | |
|---|---|---|---|---|---|
| Biosignature | Production Environment | Concentration on Earth | False positives and Secondary Outcomes | Main Spectral Features ($\mu$m) (strongest **bolded**) Via HITRAN and NIST | Citation (see references therein) |
| Organic sulfur gases e.g. DMS (($CH_3$)$_2$S) DMDS (($CH_3$)$_2$S$_2$) | marine/lacustrine settings | ~100 ppt | Secondary ethane signature may be produced abiotically, if sulfur gases do not reach detectable levels | 2.3, 3.4, ~7 (DMS, DMDS), ~9 (DMDS), ~10 (DMS, $CH_3$SH), 14 (DMS, $CH_3$SH), 18 (DMDS) | Pilcher 2003; Domagal-Goldman et al. 2011 |
| Halomethanes: Methyl Chloride ($CH_3$Cl) Methyl Bromide ($CH_3$Br) | Marine algae, terrestrial and marine algae; salt marches and wetlands | ~500 ppt ($CH_3$Cl) ~9 ppt ($CH_3$Br) | High T perchlorate pyrolysis (limited in planetary context), Exogenous delivery, Limited hydrothermal/high T processes | 3.3, 7.0, **9.9**, **13.7** (Cl) 3.5, 6.9, **10.2**, **16.4** (Br) | Segura et al. 2005; Leung et al. 2022 |
| Organic Haze | Photochemical product of $CH_4$, $CH_3$X species | Not present at high altitude on modern Earth due to oxidizing conditions | Abiotic methane can lead to haze production | NIR spectral slope, ~6 | Arney et al. 2018 |
| Phosphine ($PH_3$) | Strictly anoxic environments including paddy fields, lakes/rivers, wetlands/marshes | Ppq-ppb levels, spatially and temporally variable | Phosphite & Phosphate degradation, lightning, volcanism, exogenous delivery | 2.9, 3.4, 4.4, 5.0, 8.9, **9.5** (broad) | Sousa-Silva et al. 2020 |
| Isoprene ($C_5H_8$) | Deciduous trees and land plants | ~1-5 ppb localized and time-varying | No known (energetically unfavorable) | 3.4, 5.6, 6.3, 9.4, **10.2**, **11.4** | Zhan et al. 2021 |
| Methanol ($CH_3OH$) | Plants via demethylation of pectin | 10 ppb (surface) | Abiotic photochemistry, comets & primitive material | 2.9, 3.3, 4.6, 7.5, **9.7** | Huang et al. 2022b |
| Carbonyls e.g. Formaldehyde ($CH_2O$) | 80% of all biological compounds contain carbonyls | ~0.4 ppb (CDC) | Detectable signature is CO, which can be produced abiotically (see Schwieterman et al., 2019; Wogan & Catling 2020) | (CO) 1.18, 1.57, 2.35, **4.6** | Zhan et al. 2022 |
| Ammonia ($NH_3$) | Anerobic nitrogen fixing organisms, ammonification and nitrate reduction | Spatially and temporally variable on Earth | Volcanism, photochemistry, high temperature synthesis, equilibrium thermochemistry | 1.2, 1.3, 1.5, 2.0, 2.3, 3, 3.9, 6.1, **10.5** (broad) | Seager et al. 2013a; Huang et al. 2022a |

***Ammonia ($NH_3$).*** Nitrogen ($N_2$) is fixed to bioavailable forms like $NH_3$ by microorganisms called diazotrophs in an endothermic process (Gruber 2008). (Because nitrogen is triple bonded, this process is particularly energy-demanding.) These bioavailable forms of nitrogen are precursors to all N-bearing biological molecules, including amino acids, proteins, and nucleic acids. Diazotrophy is the source of most fixed nitrogen on Earth, with small abiotic sources from lightning (Navarro-González et al. 2001; Schumann and Huntrieser 2007). $NH_3$ can also result from the abiotic or biologically mediated degradation of organic matter, though most of this fixed nitrogen originated via biologically mediated $N_2$ fixation. $NH_3$ is volatile and enters the atmosphere in gaseous form. However, the atmospheric lifetime and concentration of $NH_3$ are low due to its high solubility and short photochemical lifetime due to photolysis and reactions with radical species (Kasting 1982). The photochemical lifetime would be longer in an $H_2$-rich atmosphere (Seager et al. 2013a; Huang et al. 2022a; Ranjan et al. 2022).

Seager et al. (2013) propose a hypothetical metabolism for $NH_3$ production in $H_2$-rich atmospheres ($N_2 + 3H_2 \rightarrow 2NH_3$), which is energy-yielding but kinetically inhibited at habitable temperatures and unknown to exist on Earth. A larger production rate and longer photochemical lifetime could result in detectable amounts of $NH_3$ in super-Earth atmospheres (Huang et al. 2022a). At sufficiently high fluxes, $NH_3$ could enter a "photochemical runaway" in analogy to $O_2$ on Earth (Ranjan et al. 2022). $NH_3$ absorbs strongly at 2.0, 2.3, 3.0, 5.5–6.5, and 9–13 μm. It would be more strongly detectable in $H_2$ atmospheres in transit transmission spectroscopy due to the extended scale height of $H_2$-rich atmospheres and, consequently, larger spectral signatures (Phillips et al. 2021). Abiotic sources of $NH_3$ include equilibrium reactions in giant planet atmospheres (Madhusudhan et al. 2016) and outgassing of highly reduced planetary interiors (Schaefer and Fegley 2010; Liggins et al. 2022). Strong kinetic disequilibrium on a temperate planet with an $H_2O$-rich troposphere would indicate biotic $NH_3$, in analogy to the logic applied for interpreting $CH_4$ biosignatures (Thompson et al. 2022). Huang et al. (2022) provide a recent comprehensive treatment of the context for $NH_3$ as an exoplanet biosignature.

***Phosphine ($PH_3$).*** Phosphorous is an essential and often limiting nutrient for life (Syverson et al. 2021). The production of phosphine ($PH_3$) by life is associated with anaerobic environments, as $PH_3$ interferes with aerobic metabolism (Nath et al. 2011), and high concentrations of $O_2$ efficiently oxidize $PH_3$. The exact mechanism for the biological production of $PH_3$ is currently unknown (and may involve indirect production via organic decay); however, it is confidently associated with life as it is produced in some microbial cultures (Jenkins et al. 2000; Liu et al. 2008) and has dynamic diurnal variability in terrestrial and marine environments, suggesting links to overall metabolic activity (Roels et al. 2005; Zhu et al. 2007). Bains et al. (2019) show that in some circumstances, phosphine could be produced by coupling bacterial phosphate reduction and phosphite disproportionation. In a planetary atmosphere, $PH_3$ is destroyed by reactions with radical species (e.g., $PH_3 + OH \rightarrow PH_2 + H_2O$) and photolysis (Cao et al. 2000; Glindemann et al. 2005). At sufficient fluxes, $PH_3$ can potentially build up to detectable levels on anoxic ($H_2$-rich or $CO_2$-rich) planets (Sousa-Silva et al. 2020; Angerhausen et al. 2023). $PH_3$ has strong infrared absorption bands at 2.9, 4.4, and 9.5 μm (the latter includes broad absorption 8-11 μm; Sousa-Silva et al. 2015; Gordon et al. 2022). Notably, $PH_3$ is produced abiotically in high-temperature layers of $H_2$-rich planetary atmospheres (Visscher et al. 2006) and could be introduced at limited rates via volcanism or micrometeorite collisions (Omran et al. 2021; Truong and Lunine 2021). As is the case for $CH_4$ and $NH_3$, $PH_3$'s interpretation as a biosignature relies on its strong kinetic disequilibrium in temperate atmospheres and the consequently high surface fluxes required to maintain a detectable concentration (Sousa-Silva et al. 2020). It has been claimed that phosphine has been detected in the Venusian atmosphere (Greaves et al. 2020, 2022); though others dispute the detection (Akins et al. 2021; Lincowski et al. 2021; Villanueva et al. 2021; Cordiner et al. 2022). This tentative $PH_3$ detection has spurred an extensive community effort to understand the potential abiotic and biological sources of $PH_3$ (e.g., Bains et al. 2021). Sousa-Silva et al. (2020) present an extensive investigation of the potential for $PH_3$ to serve as an exoplanet biosignature.

*Isoprene ($C_5H_8$).* Isoprene ($C_5H_8$) is a significant biotic volatile organic compound on Earth, produced at a rate comparable to methane (Müller et al. 2008; Zhan et al. 2021). Deciduous trees and other organisms, including bacteria and animals, generate $C_5H_8$ (Fuentes et al. 1996; King et al. 2010). In the Earth's atmosphere, isoprene undergoes rapid destruction primarily through reactions with OH radicals and subsequent reactions with $O_2$, forming various reactive products and contributing significantly to aerosol formation (Palmer 2003; Medeiros et al. 2018). Consequently, the photochemical lifetime of $C_5H_8$ in Earth's modern atmosphere is less than three hours (Zhan et al. 2021), which explains its low concentration (spatially and temporally dynamic, but ranging from ppt to the ppb levels) despite its large production rate. This photochemical lifetime would be longer in an $H_2$-rich atmosphere, allowing for greater accumulation; however, Zhan et al. (2021) find it would require a $C_5H_8$ flux ~100 times greater than Earth to be detectable. The strongest spectral features of $C_5H_8$ are at 3.4, 5.6, 6.3, 9.4, 10.2, and 11.4 μm with significant overlap between adjacent bands (Sharpe et al. 2004; Gordon et al. 2022). While it would be difficult for $C_5H_8$ to accumulate to uniquely identifiable concentrations (particularly given its absorption features overlap with other C-bearing species), it has no known abiotic false positives in a planetary context, which motivates adding this gas to our list of potential exoplanet biosignatures. Zhan et al. (2021) provide a detailed assessment of the biosignature potential of $C_5H_8$.

*Biogenic sulfur gases.* Organosulfur gases like dimethyl sulfide (DMS; $(CH_3)_2S$), dimethyl disulfide (DMDS; $(CH_3)_2S_2$), and methanethiol ($CH_3SH$) are produced by various organisms, including marine algae and photosynthetic bacteria (Visscher et al. 1991, 2003; Hu et al. 2007). Many of these organosulfur gases are indirect products of metabolism via the degradation of sulfur-containing biomolecules such as dimethyl sulfoniopropionate (DMSP; Stefels et al. 2007). However, it has recently been claimed that diverse species, including bacteria, haloarchaea, and algae, can methylate inorganic $H_2S$ into DMS as a detoxification strategy (Li et al. 2023), demonstrating these gases are sometimes direct, rather than incidental, biogenic products. Other sulfur gases such as OCS, $H_2S$, and $CS_2$ are produced by life but have robust abiotic sources, such as volcanism (Arney et al. 2014). The concentrations of these sulfur species are spatially and temporally variable (Zhang et al. 2020). Above surface ocean DMS concentrations can range up to over 100 ppt and are maintained by a global flux of about ~0.4 Tmol/year (Hulswar et al. 2022). DMS, DMDS, and $CH_3SH$ are efficiently destroyed by radical species, including OH, in Earth's $O_2$-rich atmosphere (Fung et al. 2022). The photochemical lifetimes of these species are consequently short, with a lifetime of ~1 day for DMS (Fung et al. 2022).

Pilcher (2003) suggested organosulfur gases could serve as remote biosignatures on anoxic worlds similar to the Archean Earth because photochemical lifetimes would be longer, and a reducing biosphere may be more conducive to producing these gases. Domagal-Goldman et al. (2011) investigated this scenario with a photochemical model. They find it difficult to accumulate spectrally relevant concentrations of observable DMS, DMDS, or $CH_3SH$ except for the case of a low-activity M dwarf host star. They find that $C_2H_6$ could be a secondary biosignature of a sulfur biosphere, as it is a photochemical product from the $CH_3$ radicals sourced from these gases. Importantly, Domagal-Goldman et al. (2011) investigated anoxic, $CO_2$-rich scenarios with no haze; alternative atmospheric scenarios may be more amenable to the accumulation of sulfur gases. Biogenic sulfur gases may be a compelling biosignature target on $H_2$-rich Hycean worlds and more easily detectable than on planets with high molecular weight atmospheres (Madhusudhan et al. 2021, 2023). Meadows et al. (2023) find that of all the biogenic sulfur gases, $CH_3SH$ is the most detectable in the TRAPPIST-1 planetary system, and show that a sulfagenic biosphere could produce a $CH_3SH$ signature detectable in 50 transits with JWST's NIRSpec instrument.

All three gases have strong but overlapping features throughout the IR (Sharpe et al. 2004). DMS has significant absorption features at 2.3, 3.4, 6-7, 10, and 14 μm. DMDS and $CH_3SH$ have features that are only somewhat offset from these values (see **Table 2**). Methylated organosulfur gases are not made in equilibrium reactions; however, they can be generated by electric discharge in atmospheric mixtures of $CH_4$ and $H_2S$ (Raulin and Toupance 1975). Further work is required to fully elucidate and quantify the potential abiotic false positives for methylated organosulfur gases.

**Halomethanes ($CH_3Cl$, $CH_3Br$, etc.).** Halomethanes, including $CH_3Cl$ and $CH_3Br$, are biologically produced by many organisms, including marine and terrestrial micro- and macroalgae, some plants, bacteria, and fungi (e.g., Saini et al. 1995; Manley et al. 2006; Paul and Pohnert 2011; see Table 1 in Leung et al. 2022). Halomethanes include methyl halides ($CH_3X$, where X is F, Cl, Br, or I) and polyhalomethanes with the generalized formula $CH_{4-a,b,c,d}F_aCl_bBr_cI_d$; e.g., $CH_2I_2$, $CH_2Br_2$, $CHCl_3$, $CBr_4$). Notably, halomethanes containing F are not known to be produced biologically (Leung et al. 2022). The most abundant halomethane in Earth's atmosphere is $CH_3Cl$, with a concentration of ~500 ppt (Seinfeld and Pandis 2016). $CH_3Cl$ and other halomethanes are primarily destroyed by reactions with OH radicals and photolysis (Yang et al. 2005). Segura et al. (2005) first suggested $CH_3Cl$ may be a compelling biosignature on Earth-like planets orbiting M dwarfs due to the reduced NUV radiation that would result in sharply reduced photochemical sinks and, thus, higher steady-state abundances (i.e., ~1 ppm vs. ~500 ppt for the same flux).

Subsequently, additional authors have evaluated the potential photochemical accumulation and detectability of $CH_3Cl$ on exoplanets (e.g., Rugheimer et al. 2013; Wunderlich et al. 2021). However, $CH_3Cl$ is just one of many biogenic halomethane species, most of which have not been robustly examined as potential exoplanet biosignatures.

Leung et al. (2022) provide an overview of the potential for halomethanes as a general biosignature class and include a quantitative assessment of $CH_3Cl$ and $CH_3Br$ biosignatures over a wide range of molecular fluxes and stellar host star spectral types. They find that introducing multiple halomethane species results in co-additive photochemical and spectral effects since these species compete for the same photochemical sinks and absorb at similar wavelengths. $CH_3Cl$ has its strongest features at 3.3, 7.0, 9.9, and 13.7 μm, while $CH_3Br$ has its strongest features at 3.5, 6.9, 10.2, and 16.4 μm. JWST is unlikely to detect methylated halogens because their strongest absorption features are in the MIR where signal-to-noise ratios are poor (Meadows et al. 2023). However, halomethanes may be detectable in MIR-emitted light spectra by the LIFE mission (Angerhausen et al. 2024), assuming source fluxes that are somewhat higher than Earth's global average but within the range observed in local environments. Potential abiotic sources of halomethanes include exogenous delivery of cometary material, trace levels of volcanic emission, and high-temperature perchlorate pyrolysis (Frische et al. 2006; Keppler et al. 2015; Fayolle et al. 2017), but it is unlikely these sources would approach the biogenic fluxes observed on Earth (Leung et al. 2022).

***Methylation as a general biosignature.*** Methylation is a fundamental biochemical process performed by organisms across all domains of life on Earth and often results in the formation of volatile and spectrally active products. These methylated gaseous products tend to have limited abiotic sources. Additional potential biosignature gases include—but are not limited to, methylated chalcogens, metals, and metalloids such as $(CH_3)_2Se$ (DMSe), $(CH_3)_2Se_2$ (DMDSe), $(CH_3)_2Te$ (DMTe), $(CH_3)_2Hg$, $(CH_3)_2As$, $(CH_3)_3Sb$, and $(CH_3)_3Bi$ (Fatoki 1997; Basnayake et al. 2001; Bentley and Chasteen 2002; Thayer 2002; Meyer et al. 2008; Ellwood et al. 2016; Yang et al. 2016). These species absorb strongly in the IR (e.g., Gutowsky 1949); however, kinetic rate and opacity data are incomplete, and/or the most recent measurements are highly outdated. Another challenge for these gases as biosignatures is their low global production rates on Earth. Additional work on measuring fundamental spectral and chemical inputs could be highly impactful in exploring this group of potential biosignatures.

***Additional gases.*** The list above is not exhaustive. A vast array of small molecules can be produced by life, including many heretofore unexamined in the voluminous and expanding corpus of exoplanet biosignature literature. Seager et al. (2016) propose systematically examining these small molecules for their biosignature potential. Such an effort would require an in-depth examination of their biological and abiotic sources, photochemical longevity in diverse atmospheres, and spectral features (i.e., detectability, survivability, and specificity). This would, in turn, involve studies of laboratory kinetics (reaction rates, dissociation cross-sections), fundamental spectroscopic measurements in the Vis/NIR/MIR, and end-to-end planetary simulations using atmospheric photochemistry and spectral models in addition to instrumental models for future observatories. Moreover, abiotic production due to planetary and/or astrophysical processes would have to be examined rigorously. Due to the vastness of the undertaking, our guidebook for finding life elsewhere remains incomplete, though tremendous progress has been made to enhance our knowledge of possibilities in recent years.

**Frameworks for assessing potential atmospheric biosignatures**

As discussed above, detecting any single gaseous species is likely insufficient evidence for life without adequate planetary context (Krissansen-Totton et al. 2022; Meadows et al. 2022). This context may—but not necessarily—be provided by the detection of multiple biosignature features. We also lack a complete understanding of false positives and secondary processes which may influence biosignature detection and interpretation. Therefore, many have suggested possible "agnostic" approaches to biosignatures, i.e., searches for general patterns that may indicate living processes without overreliance on our understanding of Earth life (Johnson et al. 2018; Walker et al. 2018). However, any agnostic approach to exoplanet biosignatures must grapple with the limitations of remote data. Below, we briefly summarize the logical basis for each proposed framework and refer the curious reader to studies that describe them more comprehensively.

*Thermodynamic disequilibrium and biosignature pairs.* Earlier sections discussed thermodynamic and kinetic disequilibria as applied to specific biosignature gases. Here we generalize these concepts as examples of agnostic biosignatures. The thermodynamic disequilibrium between atmospheric $O_2$ and $CH_4$ has long been considered a marker of biology's influence on Earth's atmosphere because, in equilibrium, these constituents react together to form $CO_2$ and $H_2O$ (Lovelock 1965, 1975; Hitchcock and Lovelock 1967; Sagan et al. 1993). $N_2O$ and other trace biotic species (e.g., $CH_3Cl$) would also effectively disappear from Earth's atmosphere if reacted to equilibrium. A weakness of thermodynamic disequilibrium biosignatures is that temperate terrestrial atmospheres do not foster equilibrium reactions without catalysts. For example, the reaction $2O_2 + CH_4 \rightarrow CO_2 + 2H_2O$ that underpins the $O_2 + CH_4$ biosignature pair does not occur directly under these conditions but through intermediate photochemical steps ultimately initiated by photolysis. On the other hand, the abundances of atmospheric species used to quantify thermodynamic disequilibrium are potentially observable and quantifiable, while the kinetic processes that affect them are sometimes less likely to be so.

Total atmospheric disequilibrium can be quantified by comparing the difference in Gibbs free energy between the observed and equilibrium states, equivalent to the untapped chemical energy in the planet's atmosphere, often given in joules per mole of atmosphere (Krissansen-Totton et al. 2016). The largest disequilibrium in the Earth system, and the solar system in general, is the co-existence of abundant $N_2$, $O_2$, and liquid water, much larger than Earth's $O_2$-$CH_4$ atmospheric disequilibrium (Krissansen-Totton et al. 2016). This $N_2$-$O_2$-$H_2O_{(aq)}$ disequilibrium has likely existed since the oxidation of Earth's atmosphere and was preceded by biogenic disequilibrium between $CO_2$ and $CH_4$ (Krissansen-Totton et al. 2018b). Wogan and Catling (2020) distinguish "edible" and "non-edible" disequilibria. For example, the photochemically produced disequilibrium between CO and $O_2$ in the atmosphere of Mars leaves easily exploitable free energy ("edible") on the table. This concept has been used to set an upper limit on the chemosynthetic biosphere that could exist in the Martian subsurface (Sholes et al. 2019). In contrast, Earth's $CH_4$-$O_2$ and $N_2$-$O_2$-$H_2O_{(aq)}$ disequilibria are largely kinetically inhibited (i.e., not edible), while maintained by oxygenic photosynthesis.

So long as the abundances of major gas species (or surface features in the case of an ocean) can be retrieved via spectral observations, planetary atmospheric disequilibria can be quantified. Krissansen-Totton et al. (2018a) demonstrated that $CH_4$-$CO_2$ disequilibria could be retrieved on a version of TRAPPIST-1e similar to the Archean Earth. Similarly, Young et al. (2024) showed that chemical disequilibrium biosignatures can be inferred from Proterozoic Earth-like planets with sufficient $O_2$ levels. **Table 3** provides a summary of potential biosignature disequilibrium pairs.

| Table 3. Disequilibrium Biosignature Combinations | | | |
|---|---|---|---|
| Pairing | Indication of | Observational Strategy (μm) (e.g.) | Citations: |
| $O_2$/$O_3$ + $CH_4$ (oxic atmospheres) | "redox disequilibrium", since oxidation to $CO_2$ and $H_2O$ does not occur, sources of $O_2$ and $CH_4$ must be present | Combined detections of $O_2$: 0.76, $O_3$: 9.65, $CH_4$: 1.65, 3.3, 7.7 | Lovelock 1965; Hitchcock and Lovelock 1967; Sagan et al. 1993 |
| $CO_2$ + $CH_4$ (anoxic/Archean-like atmospheres) | non-volcanic source of $CH_4$ if observed without significant CO | Combined detections of $CH_4$: 1.65, 3.3, 7.7 $CO_2$: 1.6, 2.0, 4.3, 15 | Krissansen-Totton et al. 2018a,b |
| $N_2$ + $O_2$ + $H_2O$ | largest thermodynamic equilibrium on the modern Earth | Combined detections of oceans (via glint or polarimetry), $N_2$ (via $N_2$-$N_2$ CIA centered at 4.3 μm or scattering), $O_2$: 0.76 | Krissansen-Totton et al. 2016 |
| $N_2O$ + $O_2$/$O_3$ and/or $CH_4$ | "redox disequilibrium," since oxidation states differ between $N_2O$, $O_2$, and $CH_4$ | Combined IR detection of $N_2O$: 7.8, 8.5, $O_3$: 9.65, and $CH_4$: 7.7 | Kaltenegger 2017; Schwieterman et al. 2022 |

***Kinetic disequilibrium.*** As noted above, in temperate (i.e., "habitable") planetary atmospheres, gaseous constituents are not determined by equilibrium processes but by chemical kinetics (Catling and Kasting 2017). $CH_4$ is destroyed in our $O_2$-rich atmosphere by radical species such as OH, which are downstream photolysis products dependent on UV photons from the Sun (Jacob 1999). Other gases, such as $N_2O$, are primarily destroyed directly by photolysis or can be consumed via wet or dry deposition. These sinks' relative contributions strongly depend on the planetary context, such as the background bulk atmospheric composition and the spectral characteristics of the planet's host star (Figs. 8-9). Given an understanding of these environments and a retrieved gas abundance, one can use a photochemical model to infer the production rate (molecular flux) needed to sustain that abundance. Sufficiently high production rates can fingerprint biological activity if no known abiotic source can plausibly sustain that flux. A biochemical model can further estimate the biomass needed to generate this production rate (Seager et al. 2013b). This exercise can also be inverted to preemptively assess the plausibility of potential biosignatures given the biomass needed to produce a detectable amount of that biosignature (Seager et al. 2013b). A downside of kinetic biosignatures is that they can require environmental information that may be observationally inaccessible, such as surface or subsurface chemical sinks. However, observable or otherwise inferable sinks can be used to determine a lower limit on gas production rates, which may be sufficient to infer the biogenicity of a putative biosignature gas.

*__Probabilistic approaches.__* As detailed in the previous sections, the detection and interpretation of biosignatures could be a daunting task as many biosignature gases suffer from false positives and potentially false negatives. The statistical assessment of biosignatures is an approach designed to delineate uncertainties in their detection and interpretation and assign confidence levels to the putative detection of life. Catling et al. (2018) put forward a framework for assessing exoplanet biosignatures based on Bayes' theorem and included several confidence levels for biosignature detection. Bayesian approaches to biosignatures evaluate new information conditioned on both contextual information (e.g., astrophysical and planetary properties) and prior probabilities (i.e., the likelihood of life originating in a certain environment). Walker et al. (2018) further extend this concept to interrogate how Bayesian formalism could guide searches for life on exoplanets and inform our understanding of the likelihood of the origin of life. Additional alternative probabilistic approaches include Signal Detection Theory (Pohorille and Sokolowska 2020). The difficulty in probabilistic biosignature assessment is that it relies in part on information that is hard to quantify on absolute terms, such as the probability of abiogenesis (Walker et al. 2018). Nevertheless, statistical assessments demonstrate that a single robust biosignature detection would imply biospheres are common in the universe, providing compelling motivation despite the challenge (Balbi and Grimaldi 2020). Probabilistic methods continue to be developed and refined to advance the search for life elsewhere (Bixel and Apai 2021; Madau 2023; Kipping and Wright 2024). A review of probabilistic biosignatures is available in Walker et al. (2020).

*__Network-based biosignatures.__* Atmospheric chemistry can be depicted graphically with networks, where the nodes symbolize chemical species, and the connections represent reactions (Solé and Munteanu 2004; Fisher et al. 2022). The atmospheric chemical reaction networks of planets in the solar system have distinct structures, with Earth's network standing out as the least random among planetary networks by various metrics (Holme et al. 2011; Wong et al. 2023). Indeed, Earth's atmospheric reaction network shows similarities to the topologies found in biological metabolic networks (Solé and Munteanu 2004; Fisher et al. 2022). This suggests network topologies could serve as agnostic exoplanet biosignatures (Fisher et al. 2022; Wong et al. 2023). Network-based biosignatures are an emerging area that may lead to unique insights facilitating the identification of planetary biospheres; however, additional theoretical work is required to quantify the observability of these metrics.

## SURFACE AND TEMPORAL BIOSIGNATURES

Surface and temporal biosignatures present additional avenues for the remote detection of life (Table 4). A surface biosignature is created when a component or product of living organisms, such as a pigment, introduces a detectable pattern like a spectral or polarization signature onto a planetary surface. These signatures are then recorded in reflected, transmitted, or scattered light. To accurately interpret surface reflectance signatures, one must consider not just the interaction of light with living materials (such as transmission and scattering in cellular or community structures) but also how the host star's radiation interacts with the planet's atmosphere, reaches its surface, and is then reflected or scattered to an observer. Often complementary to surface signatures, a temporal biosignature refers to changes over time in observables traceable back to a living process. This could manifest as a seasonal variation in the intensity or location of a surface biosignature or as fluctuations in a detectable gas like $CO_2$, resulting from shifts in the global balance of photosynthesis and respiration due to seasonal changes in environmental conditions (Hall et al. 1975; Meadows 2006, 2008), i.e., the seasonal component of the $CO_2$ "Keeling curve" (Keeling et al. 1996).

Surface biosignature detection requires observations of reflected light, which remain out of reach for JWST but may be obtained by HWO or a yet more advanced future observatory (e.g., Apai et al. 2019). An additional complication for current observational strategies is that temporal biosignatures require multi-epoch observations to capture changes over time, while transmission spectra of exoplanets are always obtained at the same orbital phase. Given these challenges, we offer a briefer overview of surface and temporal biosignatures than their gaseous counterparts. These types of biosignatures will likely only come into play as additional characterization of an already promising biosphere candidate, provided sufficient technological capabilities from a reflected light direct-imaging mission such as HWO (The LUVOIR Team 2019; Gaudi et al. 2020; Meadows et al. 2022). Schwieterman (2018) provides a more extensive review solely focused on potential surface and temporal biosignatures.

| | | | Observational | |
|---|---|---|---|---|
| **Table 4.** Surface & Temporal Biosignatures | | | | |
| Biosignature | Origin (e.g.) | False Positives/Unknowns | Observational Strategies | Citations |
| Photosynthetic pigments | Chlorophyll and bacteriochlorophyll pigments | Non biological pigment mimics such as minerals | Visible & NIR reflectance spectra via direct imaging | Kiang et al. 2007a, b |
| Non photosynthetic pigments | Microorganism reflectance from pigments with diverse functions | Non biological pigment mimics such as minerals | Visible & NIR reflectance spectra via direct imaging | Schwieterman et al. 2015a |
| Retinal pigments | Haloarchaea bacteriorhodopsin | Nonbiological pigment mimics such as minerals | Visible & NIR reflectance spectra via direct imaging | DasSarma and Schwieterman 2021 |
| Vegetation Red Edge | Plant leaves; phototroph cell structure | Surface materials with sharp spectral breaks | Visible & NIR reflectance spectra via direct imaging | Seager et al. 2005 |
| Chirality | Nearly all biological molecules on the Earth have L-amino acids and D-sugars | Unclear how universal this characteristic is, enantiomeric excess also seen in carbonaceous chondrites. Weak signal | Can be revealed via spectropolarimetry although signal is weak | Sterzik et al. 2012; Patty et al. 2018; Sparks et al. 2021; Gleiser 2022 |
| Florescence & Bioluminescence | Reprocessing or direct production of light from biological organisms | Requires very high signal to noise and probably low cloud cover; non biological processes can also fluoresce; bioluminescence produces very weak signatures | Reflectance spectra via direct imaging, peaks in visual wavelengths | Papageorgiou 2007; Haddock et al. 2010; O'Malley-James and Kaltenegger 2018 |
| Seasonality | Temporal changes in gas levels due to productivity changes as the planet orbits | Requires significant time investment and retrieval accuracy | Time series retrievals of abundance of gases linked to biological outputs or inputs, such as $O_3$ or $CO_2$ over time | Olson et al. 2018; Mettler et al. 2023 |

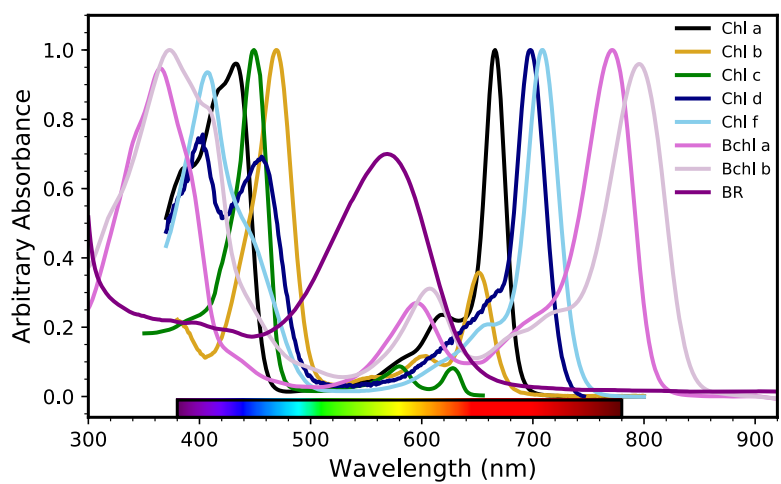

**Figure 11.** Chlorophyll (Chl), bacteriochlorophyll (Bchl), and bacteriorhodopsin (BR) absorbance spectra. The figure was taken from DasSarma and Schwieterman (2021) under .

### Surface signatures of photosynthesis

Most of Earth's biomass is produced via photosynthesis, so it stands to reason that photosynthetic biomass may afford the most extensive surface signature of life on an inhabited planet. We offered a brief review of oxygenic photosynthesis in an earlier section and here consider the photosynthetic pigments and resulting surface biosignatures. Light absorption is facilitated by pigments called chlorophylls in oxygenic phototrophs (e.g., plants, algae, cyanobacteria) Chlorophyll *a* (Chl *a*) and *b* (Chl *b*) are common in vegetation, and absorb light in the red and blue wavelengths, with peaks at 435 and 660-670 nm for Chl *a*, and 460 and 650 nm for Chl *b* (see Table 1 in Schwieterman et al., 2018 for absorption maxima of all photosynthetic pigments). The green color of chlorophylls arises from weaker absorption in the green spectrum and reflection of this color to the observer.

However, photosynthesis can also occur without producing oxygen in a variety of environmental settings. This process is more evolutionarily ancient than oxygenic photosynthesis and is called anoxygenic photosynthesis (Olson 2006). It can be represented by the following general formula:

$$CO_2(g) + 2H_2A^w + h\nu \rightarrow (CH_2O)_{organic} + H_2O + 2A$$

where $CO_2$ is carbon dioxide gas, $H_2A$ is an electron source (such as $H_2S$ or $H_2$, $h\nu$ is light energy, $(CH_2O)_{organic}$ represents biomass, and $2A$ is an oxidized waste product.

Photosynthesis is the conversion of light energy into chemical energy and involves a complex series of oxidation/reduction reactions. Light is absorbed by photosynthetic pigments in both oxygenic and anoxygenic phototrophs, and electrons in special types of pigments called reaction centers are excited and donated to receptor molecules. Those electrons are ultimately captured and stored in energy-rich compounds such as ATP and NADPH, which are in turn used to reduce $CO_2$ to $CH_2O$. The electron donated by the pigment is replaced by the external reductant $H_2A$ or in the case of oxygenic photosynthesis, $H_2O$ (Blankenship 2002).

Anoxygenic phototrophs use special types of pigments called bacteriochlorophyll (Bchls), that absorb light primarily in the near-infrared (see Table 1 in Schwieterman et al., 2018). For example, Bchl *a* in purple bacteria has two characteristic absorption maxima at 790-810 nm and 830-920 nm. The pigment that absorbs furthest into the NIR is Bchl *b* at 1015-1040 nm, which near the theoretical limit of ~1100 nm for photon absorption in electronic transitions (Kiang et al. 2014). Phototrophs also employ accessory pigments, like carotenoids, to absorb light at shorter wavelengths and that energy is transferred to the reaction center, allowing effective harvesting of light across the visible spectrum (Allakhverdiev et al. 2016). The absorbance spectra of chlorophyll and bacteriochlorophyll pigments are detailed in Figure 11.

**The vegetation red edge (VRE) and other photosynthetic "edge" signatures**

The VRE is a possible exoplanet biosignature (Seager et al. 2005) detectable in the reflectance spectrum of a planet, and is identified by a sharp increase in reflectance at the boundary between visible and near-infrared wavelengths (~700 nm), most strongly seen in green vascular plants today (though not limited to them). This effect arises from the contrast in chlorophyll absorption in the red (~650 – 700 nm) and scattering properties of cellular structures in the near-infrared (~750-1100 nm) (Gates et al. 1965; Knipling 1970). The VRE is evident in various oxygenic phototrophs, including plants, algae, and cyanobacteria, with wavelengths clustered between 690-730 nm, and is stronger than the chlorophyll "green bump" near 550 nm (Kiang et al. 2007b). Figure 12 illustrates the VRE along with common surface materials, such as snow and soil, other biogenic materials, including a bacterial mat. Figure 12 also shows slope change features from materials such as cinnabar and sulfur, which may constitute false positives for biogenic "edge" signatures (Seager et al. 2005).

The VRE's spectral break is commonly used in the Earth sciences to assess vegetation health and coverage of the planet through the normalized difference vegetation index (NDVI; Tucker 1979; Tucker et al. 2005). The NDVI is calculated as:

$$NDVI = \frac{NIR - VIS}{NIR + VIS}$$

where NIR is near-infrared albedo and VIS is visible band albedo. NDVI values range from -1 to 1, with higher values indicating vegetation. However, using a broad metric like NDVI for exoplanet biosignature analysis has limitations, as it struggles to distinguish between the red-edge and mineral slopes. Minerals such as cinnabar and elemental sulfur have steep slopes from visible to red wavelengths that can mimic vegetation signals (Fig. 12). Studies suggest that while NDVI scores the lunar surface similarly to Earth's vegetation, the VRE could aid in the initial identification of Earth-like worlds when combined with other data (Livengood et al. 2011).

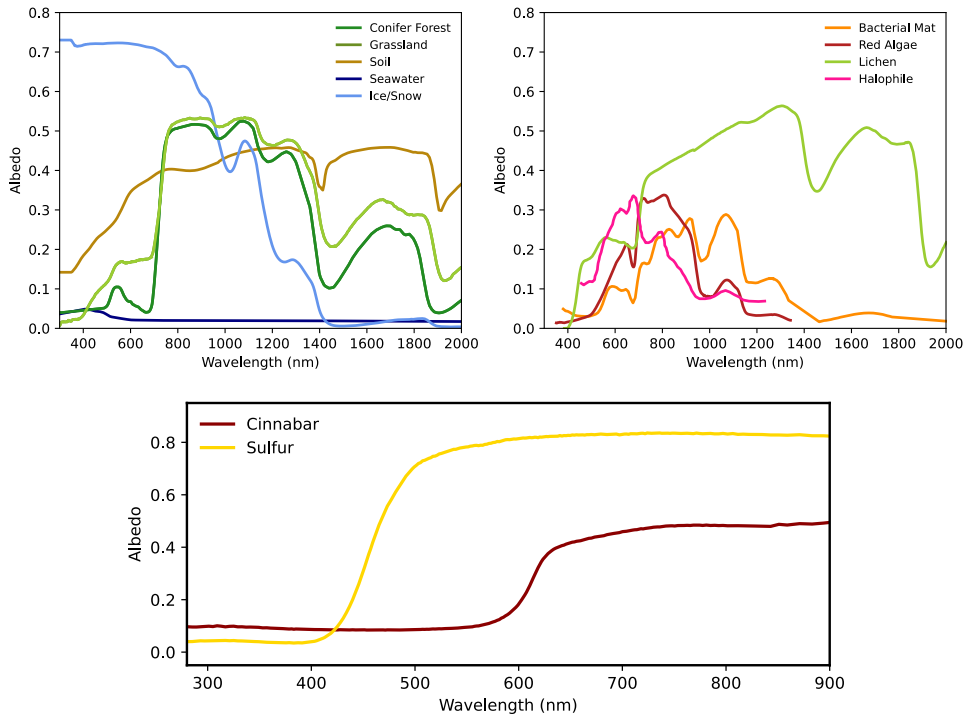

**Figure 12.** Spectral albedos of various living and non-living materials. Sulfur and cinnabar may be "false positives" for edge-like biosignatures (bottom panel) but lack the hydration bands present in the living material. The halophile spectrum is from Dalton et al. (2009), the conifer forest spectrum is sourced from the ASTER spectral library (Baldridge et al. 2009), and all other spectral albedos are sourced from the USGS spectral library (Clark et al. 2007).

The strength of the VRE likely varied significantly throughout Earth's history. Land plants responsible for the modern VRE have only existed for about 400 million years (Kenrick and Crane 1997). Algae may have supported VRE signatures for up to 1.0 – 1.6 Ga (Brocks et al. 2023). Before that, photosynthetic life was bacterial, dating back to at least 3.5 Ga (Stüeken et al. 2024). The reflectance spectra of these early surface microbial communities  would have differed from vegetation and would be likely dominated by microbial mats or stromatolites in marine intertidal and continental settings. Sanromá et al. (2013) determined that oxygenic cyanobacteria in microbial mats, if covering 50% of early Earth's land, could be distinguishable from vegetation due to dimmer reflectance at wavelengths beyond ~1.3 μm.  Furthermore, Sanromá et al. (2014) observed that mats of anoxygenic phototrophs, like purple bacteria, would show a red-shifted VRE effect due to the longer wavelength absorption of bacteriochlorophylls, with NIR edges beyond around 1 μm. However, significant coverage would be necessary for detection. Tinetti et al. (2006) found that NIR analogs to the VRE may be more easily detected in reflected light through the clouds and atmosphere. Future missions such as HWO could search for photosynthetic edge-like features analogous to the VRE on exoplanets in reflected light, which could be detected given sufficient surface coverage (Gomez Barrientos et al. 2023).

**Rhodopsin-based phototrophy**

In the search for extraterrestrial life, pigments other than those used in photosynthesis, such as bacteriorhodopsin produced by haloarchaea, are gaining attention as potential biosignatures (DasSarma 2006; DasSarma et al. 2020). Unlike chlorophyll, bacteriorhodopsin is not linked to the production of organic matter. It is a light-driven proton pump that generates cellular energy, absorbing light in the green wavelength range of the visible spectrum (~570 nm), complementary to chlorophylls and bacteriochlorophylls (Fig. 11). DasSarma and Schwieterman (2021) speculated that this spectral complementarity may reveal the timing of rhodopsin's evolutionary origins with respect to chlorophyll pigments and offer a possible alternative surface biosignature to search for on exoplanets. Sephus et al. (2022) further explored this concept, analyzing the evolutionary history of microbial rhodopsins. Initially functioning as light-driven proton pumps and primarily absorbing green light, these pigments likely evolved in response to Earth's early photic environments. This indicates that inhabited planets could show diverse spectral signatures based on their dominant pigments, which could be centered on rhodopsin-based phototrophy (DasSarma et al. 2020). Figure 13 shows hypothetical planetary spectra, contrasting those dominated by VRE-producing forests with a surface dominated by salt ponds hosting pigmented halophilic archaea with rhodopsin and carotenoid pigments. This example is for illustrative purposes, as the surface coverage fraction of these biogenic materials is unlikely to be close to 100%.

**Predicting photosynthetic surface signatures around other stars**

A star's spectral energy distribution (SED) varies with its temperature, influencing the wavelength of pigment absorption in photosynthesis. Kiang et al. (2007a) suggest that the evolution of pigment absorption might adapt to the most efficient wavelength based on the star's temperature, resulting in predictable 'edge' wavelengths. This shift is linked more strongly to surface photon irradiance than energy irradiance, considering photosynthesis requires specific photon amounts. For instance, planets orbiting cooler stars (like M-dwarfs) might exhibit biospheres with NIR-shifted red edges, while those around hotter stars (like F-dwarfs) could have blue-shifted edges. Several workers have developed theoretical models to predict photosynthetic absorption based on the SED of the host star (Lingam et al. 2021; Covone et al. 2021; Lehmer et al. 2021; Marcos-Arenal et al. 2022; Battistuzzi et al. 2023; Illner et al. 2023). However, the adaptation of 'edge' wavelengths might not directly correspond to the host star's spectrum or the wavelengths of light at the surface of the planet. This is because phototrophs, at least on Earth, are usually not limited by photon availability but rather by resources like water, electron donors, nitrogen, iron, or phosphorus (Bristow et al. 2017). Additionally, continental and marine environments offer a wide variety of light environments for which phototrophs are adapted (e.g., Kurashov et al. 2019), which are only partly influenced by the star. As such, the relationship between star spectrum and photosynthetic pigment evolution is not necessarily straightforward or fully understood.

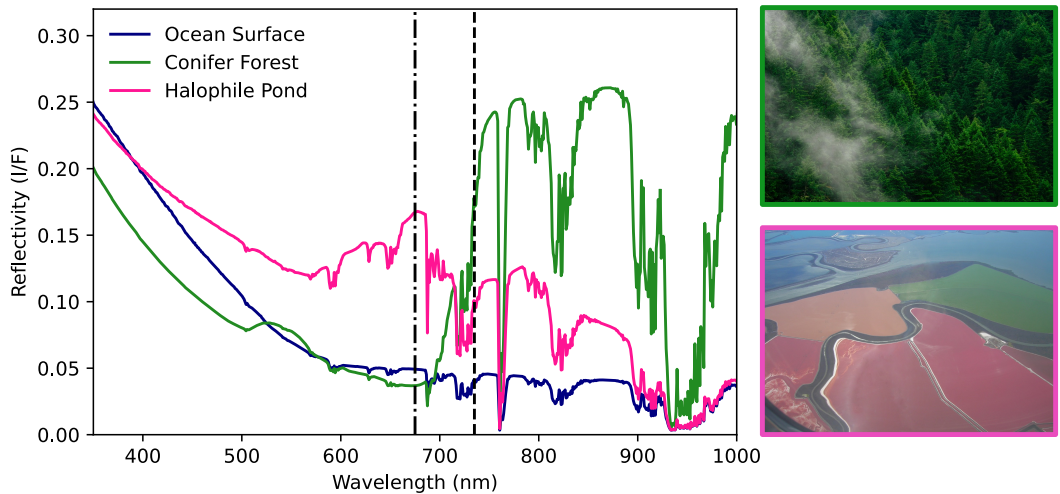

**Figure 13.** Differentiating between the surface biosignatures of vegetation and halophiles. **Left:** This panel illustrates simulated reflection spectra of planets similar to Earth, with surfaces predominantly covered by oceans, forests, or pigmented halophiles. A **dot–dashed line** at 675 nm indicates a distinct peak in the reflectance of halophile ponds. The Vegetation Red Edge (VRE) characteristic of forests is observer 700 and 750 nm (**dashed line**). The forest surface's albedo was calculated using data from the ASTER spectral library (Baldridge et al. 2009), while the halophile surface data was sourced from the spectral analysis of San Francisco Bay's salt ponds (Dalton et al. 2009). **Right:** The images display a conifer forest (**top**, from Andrew Coehlo under an Unsplash Licence) and salt ponds dominated by halophiles in San Francisco Bay (**bottom**, from Grombo, Wikipedia Commons).

### Non-photosynthetic and other surface biosignatures

Pigments can play a variety of crucial roles in organisms beyond light capture for photosynthesis, which can sometimes be completely decoupled from the light environment (Schwieterman et al. 2015a). These functions include protection against ultraviolet radiation, vital for species in environments with high UV exposure and contributions to temperature regulation by reflecting sunlight or aiding in heat absorption. Additionally, pigments are involved in visual signaling, playing a role in processes like camouflage, mating, and warning signals. Moreover, some pigments have antioxidative properties, protecting cells from damage caused by oxidative stress. Many pigments display edge features analogous to the edges seen in photosynthetic organisms and efforts have been undertaken to categorize the wide diversity of these features (Hegde et al. 2015; Coelho et al. 2022). A large and robust chemosynthetic biosphere could potentially be identified via non-photosynthetic pigments. However, just as is the case with photosynthetic signatures, the surface coverage would have to be sufficient to allow for remote detectability in planetary disk-averaged spectra.

## Polarization and Chirality

Polarization measurements, both linear and circular, have been proposed as surface biosignatures within and outside the solar system (Sparks et al. 2009b, 2012; Berdyugina et al. 2016; Patty et al. 2017; Gleiser 2022). Polarization helps discern Earth-like atmospheres and is influenced by pigments like chlorophyll (Sterzik et al. 2012, 2014, 2019). Studies show linear polarization peaks where pigment absorption is highest, contrasting with abiotic materials like sand or rock (Berdyugina et al. 2016). Circular polarization signatures can arise from chiral biological molecules. Chirality (mirror image biomolecules that can't be superimposed), and specifically homochirality (the exclusive use of one "handedness" of a biomolecule over the other), is considered a universal agnostic biosignature and is a highly specific indicator of living material (Sparks et al. 2009a, 2021; Patty et al. 2018;  Gleiser 2022; though see Avnir 2020[1] for a qualification of this view). However, circular polarization features related to chiral centers in photosynthetic pigments and interactions between pigments are weaker than linear polarization features and challenging to link definitively to surface microbial communities or vegetation in remote observations. Additional work is underway to comprehensively examine polarization signatures on Earth and their possible manifestation elsewhere in the universe (Patty et al. 2019, 2022; Gordon et al. 2023).

## Fluorescence and bioluminescence

Vegetation and microorganisms can emit light through fluorescence and bioluminescence (Papageorgiou 2007; Haddock et al. 2010). Chlorophyll autofluorescence, occurring when the pigments are excited by UV light and the absorbed photons are re-emitted at lower energy levels, is observable via Earth-orbiting satellites.  The main spectral range for chlorophyll autofluorescence is in the red 640-800 nm, with characteristic peaks at 685 and 740 nm, though the overall signals are small and require high-resolution spectroscopy to observe (Joiner et al. 2011; Sun et al. 2017). Some authors have proposed biological autofluorescence could be a temporal surface biosignature, with fluorescence from organisms responding to stellar flare activity (O'Malley-James and Kaltenegger 2018, 2019).  Komatsu et al. (2023) quantitatively examine the potential of detecting fluorescence biosignatures on exoplanets in reflected light. They find bacteriochlorophyll autofluorescence emission at 1000–1100 nm is plausibly detectable (given sufficient surface coverage of fluorescing vegetation) on planets orbiting ultracool dwarfs with an HWO-like telescope due to the coincidence of fluorescence wavelengths with atomic and molecular absorption in the stellar spectrum. Potential false positives include minerals such as fluorite and calcite, which can also be remotely detected on Earth from space (Köhler et al. 2021), though their fluorescence profiles may vary from biological fluorescence.

Complementary to florescence, bioluminescence is the active production (not reprocessing) of light by organisms. Known life accomplishes this through the oxidation of a class of molecules called luciferins and are found in diverse forms of life encompassing bacteria, single cell eukaryotes, and animals (Haddock et al. 2010). Bioluminescence of some marine microbes can be observed by Earth-observing satellites and patches of luminescent plankton can encompass more than 10,000 $km^2$ (Miller et al. 2005), leading to its suggestion as a potential future exoplanet biosignature (Seager et al. 2012). However, the overall luminosity of these luminous patches is low, and no study has thus far quantitatively examined its detectability on exoplanets.

## Seasonal biosignatures

Earth's atmosphere displays seasonal variations in gases like $CO_2$, $O_2$, and $CH_4$, which are influenced by biological processes. $CO_2$ levels fluctuate due to plant growth and decay, decreasing in spring and summer and increasing in fall and winter (Hall et al. 1975; Keeling et al. 1996). $O_2$ variation is closely related to $CO_2$ changes, influenced by photosynthesis and organic matter decay (i.e., $CO_2 + H_2O \leftrightarrow (CH_2O)_{org} + O_2$). While $O_2$'s absolute variability is greater than $CO_2$'s (Keeling and Shertz 1992), partly due to $CO_2$'s higher solubility in ocean water, its proportional change relative to background concentrations is much smaller. $CH_4$ shows a more complex seasonal pattern, with lowest levels in northern summer, a smaller dip in winter, and peaks in late fall and early spring (Rasmussen and Khalil 1981). $CH_4$'s primary non-human source is methanogenic microbes in wetlands, but its seasonal changes are largely governed by interactions with OH, mainly derived from tropospheric water and thus varying with surface temperature (Khalil and Rasmussen 1983).

The seasonal oscillations in Earth's atmospheric gases vary by hemisphere and latitude; $CO_2$ changes are more pronounced in the northern hemisphere due to greater land coverage (Keeling et al. 1996). These patterns suggest that detecting seasonal cycles on exoplanets will depend on factors like viewing angle and planet tilt. Planets with greater obliquity or eccentricity than Earth might have more detectable seasonal signatures, though one must consider viewing geometry and mixing hemispheres in a disk average. Even a maximum $CO_2$ variation of 1-3% would be difficult to detect, given weak or saturated $CO_2$ bands and competing effects from clouds and changes to atmospheric temperature structure (Mettler et al. 2023).

Olson et al. (2018) provide a detailed review of the general phenomenology underlying atmospheric seasonality as a potential exoplanet biosignature and the opportunities and challenges associated with observing it. They describe the possible impacts of obliquity, eccentricity, and viewing geometry on the observed signal. These authors proposed that $O_2$ seasonality may have been magnified on Proterozoic Earth, due to the smaller background $O_2$ concentrations and therefore larger potential for proportional changes (cf. Wogan et al. 2022). This $O_2$ would not be directly observable but could be inferred by seasonality in the UV signatures of $O_3$, which can be observed at lower $O_2$ abundances than directly (cf. Kozakis et al. 2022).

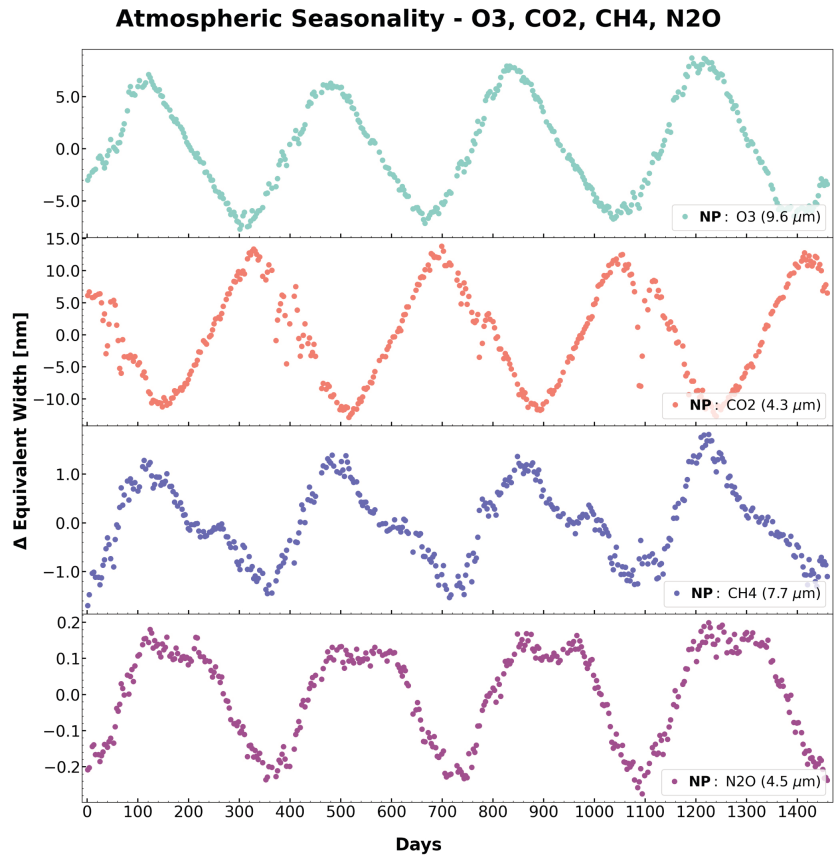

**Figure 14.** Variations in the Equivalent Width (EW; in nm) of IR spectral bands of $O_3$, $CO_2$, $CH_4$, and $N_2O$ from reconstructed disk-averaged thermal IR data from the Atmospheric Infrared Sounder (AIRS) onboard the Earth-observing Aqua satellite. The data span four years and are based on a "North Pole" view. Additional details are available in Mettler et al. (2023). This figure was sourced from Mettler et al. (2023) under <u>Creative Commons Attribution License CC-BY</u>.

Mettler et al. (2023) examine the seasonality of atmospheric gases $N_2O$, $CH_4$, $O_3$, and $CO_2$ in the IR using Earth-observing data. Figure 14 shows the equivalent width (EW) variation in IR $O_3$, $CO_2$, $CH_4$, and $N_2O$ bands at 9.6, 4.3, 7.7, and 4.5 μm, respectively, from a "north pole on" viewing geometry (Mettler et al. 2023). The spectral variations shown are small and impacted by cloud formation, changes in atmospheric temperature structure, and other seasonal changes in atmospheric and surface conditions. Nonetheless, these data demonstrate that variations in the spectral signatures of gases reveal seasonal processes, in principle. Ultimately, the observed seasonality in these gases depends on these complex factors and viewing geometry, in addition to their intrinsic changes in abundance.

Seasonal biosignatures need not be limited to gases, as the surface pigmentation of vegetation also changes seasonally, though such a biosignature would have compounding challenges. Nevertheless, seasonal biosignatures may confirm candidate biospheres once identified, perhaps in the distant future.

## CONCLUSIONS

Earth's oxygenic photosynthetic biosphere has left an indelible mark on the remote spectrum of our planet. If other planets in our local universe are inhabited, their biospheres could similarly alter the chemical composition of their planetary atmospheres and their resulting remotely detectable spectral signatures. Here, we have provided an overview of potential exoplanet biosignatures, including their sources and sinks, observable features, and possible false positives. Confirming that potential biosignatures fingerprint real biospheres requires planetary context, and no single signature is likely sufficient on its own for one to claim the presence of life. We have given particular emphasis to atmospheric (gaseous) biosignatures, as it is becoming feasible to search for them in the near to intermediate term. Potential gaseous biosignatures include those that dominate on Earth today ($O_2$, $O_3$, $CH_4$, and $N_2O$) and those known to be produced by life but with small concentrations in Earth's modern atmosphere (e.g., $(CH_3)_2S$, $CH_3Cl$, $PH_3$, $C_5H_8$), though these could accumulate to higher levels in other planetary environments. The changing oxygenation of Earth's atmosphere over geologic time likely influenced the concentrations of biogenic gases, especially $CH_4$ and $N_2O$. Thermodynamic and kinetic disequilibria can be useful metrics for evaluating potential atmospheric biosignatures. Additionally, surface biosignatures, such as the vegetation red edge, and temporal biosignatures, such as atmospheric seasonality, may offer independent lines of evidence supporting the identification of putative biospheres, though they are less easily characterized. Bayesian frameworks for biosignatures may potentially provide statistical robustness to this search and the development of further frameworks for characterizing biosignatures continues. The search for life beyond the solar system is a compelling and monumental undertaking, and we are privileged to live in a time when that search is becoming possible.

### Acknowledgments
ES and ML gratefully acknowledge support from the NASA Exobiology Program under grant no. 80NSSC20K1437, the NASA Exoplanet Research Program under grant no. 80NSSC22K0235, the NASA Interdisciplinary Consortium for Astrobiology Research (ICAR) under grant nos. 80NSSC21K0594, 80NSSC23K1399, and 80NSSC23K1398, and The Kavli Foundation. We thank Giada Arney, Ravi Kopparapu, and Niki Parenteau for their helpful comments and suggestions that improved this chapter.